\documentclass{aa} 
\usepackage[dvipsnames]{xcolor} 
\usepackage{graphicx, subfigure, amsmath, pifont}
\usepackage[breaklinks=true]{hyperref} 
\hypersetup{colorlinks=true,citecolor=Blue}
\usepackage{natbib}
\bibpunct{(}{)}{;}{a}{}{,} % to follow the A&A style
\newcommand{\cmark}{\ding{51}}%
\newcommand{\xmark}{\ding{55}}%
\usepackage[english]{babel}
\usepackage{blindtext}
\usepackage{booktabs} % nice tables
\usepackage{comment}
\usepackage{commath}
\usepackage{cancel}
\usepackage{pifont}
\usepackage{siunitx}
\usepackage[normalem]{ulem}

\newcommand{\vfrag}{v_\mathrm{f}} % fragmentation speed
\newcommand{\St}{\mathrm{St}} % Stokes number
\newcommand{\vkep}{v_\mathrm{K}} % Kepler speed
\newcommand{\cs}{c_\mathrm{s}} % sound speed
\newcommand{\absdlnPdlnr}{\left\vert\frac{\partial\ln{P}}{\partial\ln{r}}\right\vert} % absolute value of log pressure gradient
\newcommand{\hg}{h_\mathrm{g}} % gas pressure scale height
\newcommand{\sigmad}{\Sigma_\mathrm{d}} % particle column density
\newcommand{\sigmag}{\Sigma_\mathrm{g}} % particle column density
\newcommand{\DiffZ}{\delta_{\mathrm{z}}} % vertical diffusion parameter
\newcommand{\rhoint}{\rho_\mathrm{s}} % internal (solid) density of grains
\newcommand{\relvel}{\Delta v} % relative velocity
\newcommand{\tgrow}{\tau_\mathrm{grow}} % growth timescale
\newcommand{\tdrift}{\tau_\mathrm{drift}} % drift timescale
\newcommand{\vdrift}{v_\mathrm{drift}} % radial drift velocity
\newcommand{\TurbPara}{\delta_\mathrm{t}} % turbulence strength parameter
\newcommand{\Sigdtog}{Z} % dust to gas ratio of column densities
\renewcommand{\Re}{\mathrm{Re}} % Reynolds number

\begin{document} 

 \title{Growing and trapping pebbles with fragile collisions of particles in protoplanetary disks}

   \author{
   Paola Pinilla\inst{1}, 
   Christian T. Lenz\inst{1},
   and 
   Sebastian M. Stammler\inst{2}
   }
   \institute{Max-Planck-Institut f\"{u}r Astronomie, K\"{o}nigstuhl 17, 69117, Heidelberg, Germany, \email{pinilla@mpia.de}, \email{lenz@mpia.de}
   \and
             University Observatory, Faculty of Physics, Ludwig-Maximilians-Universit{\"a}t M{\"u}nchen, Scheinerstr. 1, D-81679 Munich, Germany
   }
   \date{}

  \abstract
   {Recent laboratory experiments indicate  that destructive collisions of icy dust particles occur with much lower velocities than previously thought. These fragmentation velocities play a crucial role in planet formation because they set the maximum grain size in collisional growth models. When these new velocities are considered from laboratory experiments in dust evolution models, a growth to pebble sizes (typically millimeter- to decimeter-sized particles) in protoplanetary disks is difficult. This may contradict (sub-) millimeter observations and challenge the formation of planetesimals and planets. We investigate the conditions that are required in dust evolution models for growing and trapping pebbles in protoplanetary disks when the fragmentation speed is 1\,m\,s$^{-1}$ in the entire disk. In particular, we distinguish the parameters controlling the effects of turbulent velocities ($\TurbPara$), vertical stirring ($\delta_z$), radial diffusion ($\delta_r$), and gas viscous evolution ($\alpha$), always assuming that particles cannot diffuse faster (radially or vertically) than the gas (i.e., $\delta_{r,z,\mathrm{t}}\leq\alpha$). We compare our models with observations of protoplanetary disks at both the near-infrared and millimeter regimes. To form pebbles and produce effective particle trapping, the parameter that controls the particle turbulent velocities must be small ($\delta_t\lesssim10^{-4}$). In these cases, the vertical settling can limit the formation of pebbles, which also prevents particle trapping. Therefore the parameter that sets the vertical settling and stirring of the grains must be $\delta_z<10^{-3}$. Our results suggest that different combinations of the particle  and gas diffusion parameters can lead to a large diversity of millimeter fluxes and dust-disk radii. When pebble formation occurs and trapping is efficient, gaps and rings have higher contrast at millimeter emission than in the near-infrared. In the case of inefficient trapping, structures are also formed at the two wavelengths, producing deeper and wider gaps in the near-infrared. Our results highlight the importance of obtaining observational constraints of gas and particle diffusion parameters and the properties of gaps at short and long wavelengths to better understand basic features of protoplanetary disks and the origin of the structures that are observed in these objects.}


   \titlerunning{Growth and trapping of pebbles with fragile collisions}
   \maketitle

%
%________________________________________________________________
%%%%%%%%%%
\section{Introduction}                  \label{sect:intro}
%%%%%%%%%%

Planet formation occurs in protoplanetary disks, which are mainly composed by gas. Most of the information that we have about planets forming in disks comes from the dust that dominates the disk opacity. The initial properties of the dust in disks are inherited from the interstellar medium, such as its size and composition. These initial properties are expected to change by different physical processes that lead to collisions and transport of the particles \citep[see, e.g.,][for a review]{birnstiel2016}. 

%%%%%%%%%
%% FIGURE % %
%%%%%%%%%
\begin{figure*}
 \centering
        \includegraphics[width=18cm]{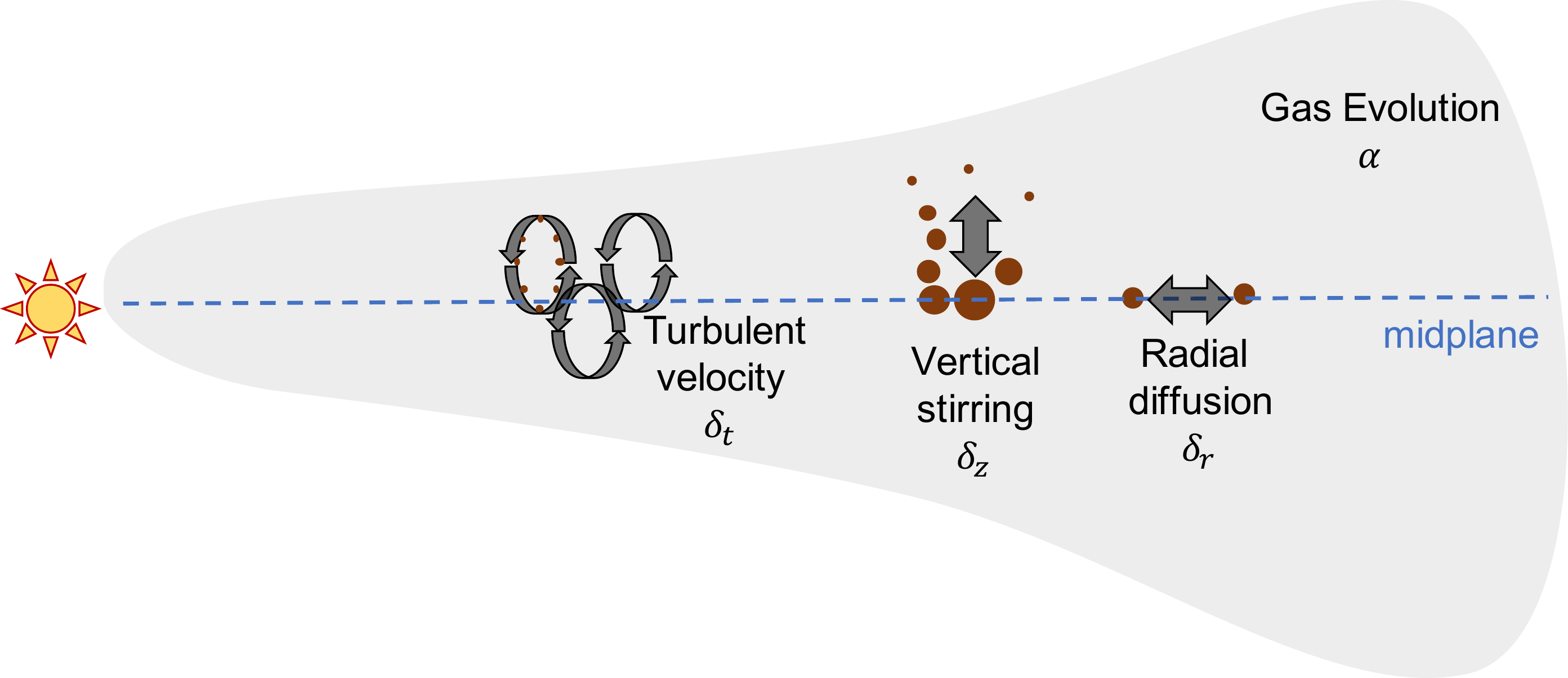}
   \caption{Schematic view of turbulent velocities, vertical stirring and settling, and radial diffusion of dust. In the standard models, they are all set to a  value that depends on the $\alpha$ parameter that regulates the viscous evolution of the gas. We assumed all four parameters to be independent.}
   \label{fig:schematic_models}
\end{figure*}

The growth of particles from (sub-)micron-sized dust to pebbles is fundamental for understanding the first steps of planet formation. From the observational point of view,  we can have access to this process by observing the distribution of micron-sized dust to pebbles (submillimeter- to centimeter-sized particles) with observations from the optical to centimeter wavelengths, respectively. Solid bodies larger than centimeter-sized particles have very low opacities, and we are blind from the observational point of view about what occurs from the formation of pebbles to planets. To connect current theories of planet formation with observations, we rely on connecting our observational knowledge of the small bodies (up to pebbles) to exoplanet statistics.

Recent deep observations of protoplanetary disks with high angular resolution have shown structures in these objects, such as rings, gaps, spiral arms, and lopsided shapes \citep[e.g.,][]{andrews2020}. These structures are believed to be regions of high pressure where the radial drift of pebbles is reduced or stopped completely \citep{whipple1972}, and where planetesimal formation can occur \citep[e.g.,][]{pinilla2012, pinilla2020, dullemond2018, lenz2019, stammler2019, klahr2020}. The current understanding of these structures from the observational and theoretical point of view is driving the field of planet-forming disks.

Another crucial part that helps to understand the first growth from interstellar medium dust to pebbles are laboratory experiments of particle collisions. They provide the potential outputs after collisions with a given initial relative velocity. Potential outputs include sticking, fragmentation, bouncing, or mass transfer \citep[e.g.,][]{windmark2012}. They also define the limits for theoreticians of velocity thresholds for effective growth or destructive collisions \citep[e.g.,][]{blum2000, blum2008, kimura2015, kimura2020}. These experiments are challenging, however, because the conditions must be set to be as close as possible to the physical properties of disks, which means low densities, pressure, and temperatures. 

It has commonly been thought that water-ice particles are more sticky than silicates \citep{blum2000, wada2009, wada2011, gundlach2015, musiolik2016}, and the comparison has been made that this would be like snowballs versus sand. The difference in the fragmentation velocity between these two species is at least one order of magnitude. High fragmentation velocities for icy particles helps greatly in forming pebbles in the disk because the maximum grain size of pebbles in the fragmentation-dominated regime is proportional to the square of the  threshold of the fragmentation velocity \citep[e.g.,][]{brauer2008, birnstiel2010}. 
Nonetheless, recent laboratory experiments have reproduced the temperatures of protoplanetary disks better, which changed our perspective on collisional growth because it was concluded that amorphous water-ice particles are as fragile as silicates or even more so \citep{gundlach2018, musiolik2019, steinpilz2019}. The purpose of these experiments is to measure the sticking force when two particles are in contact. This force depends on the surface energy of the grains and therefore on the particle composition \citep[e.g.,][]{kimura2015}. Most of the classical experiments used monolayer dust particles with an insignificant amount of water or no water at all. This limitation comes from keeping low values of temperature and pressure, and most of the information that we previously had about collision of (water) ices came from numerical experiments. More recent laboratory experiments were able to retain low temperatures in their collisions ($<$180\,K) and showed that the surface energy of these grains decreases. This contradicts previous assumptions of collisions of water-ice particles.

The results of these new experiments suggest that the fragmentation velocities for icy particles is 1\,m\,s$^{-1}$ or even lower. This poses a problem not only for the formation of pebbles and hence for the formation of planets as a whole, but also makes the current interpretation of disk observations challenging. 
If the current structures are the byproduct of pressure bumps trapping the particles, grains have to grow at least to submillimeter sizes to efficiently slow down their drift at the pressure bump and be trapped. If particles remain small  (typically $\lesssim0.1\,$mm), radial drift is slow, and other effects such as turbulent mixing and radial diffusion may become more important, preventing the grains from being trapped. The maximum grain size that particles may reach with fragmentation velocities of 1\,m\,s$^{-1}$ may agree with the maximum grain size that \cite{liu2019} and \cite{zhu2019} proposed to explain the optical depth of several disks recently observed with the Atacama Large Millimeter/submillimeter Array (ALMA) under self-scattering.

In this paper, we investigate the conditions that are required in dust evolution models to grow and trap pebbles in protoplanetary disks when the fragmentation speed is 1\,m\,s$^{-1}$. The main questions that we consider are what values of turbulence, settling, and diffusion are needed to grow dust to pebbles and have particle trapping if the
fragmentation velocities are as suggested from laboratory experiments. We define a reasonable value for the parameter that controls the viscous evolution of the gas versus dust evolution when
we compare this with the observed properties of protoplanetary disks.   To answer these questions, we investigate parameters that rule the turbulent velocities ($\TurbPara$), vertical stirring ($\delta_z$), radial diffusion ($\delta_z$), and viscous gas evolution ($\alpha$), and consider that they are independent of each other. We distinguish the effect of each of these physical processes by always assuming that particles should not diffuse faster than the gas (i.e., $\delta_{r,z,\mathrm{t}}\leq\alpha$). To test our models, we combine the results with radiative transfer simulations  to compare them with observations at near-infrared and millimeter emission. 

This paper is organized as follows. In the next section, we briefly describe the models and the  parameters that we assumed for the angular momentum transport of gas and dust (vertical and radial) diffusion. In this section we also describe our assumptions for the radiative transfer calculations. In Sect.~\ref{sect:results} we present the results of our models. In Sect.~\ref{sect:discussion} we discuss our results in the context of different observational diagnostics, such as millimeter fluxes, dust-disk radii, images, and spectral indices. Section~\ref{sect:conclusions} summarizes our conclusions.

\section{Models}
 
\subsection{Dust evolution models}
 
We used the \texttt{dustpy} code developed by Stammler \& Birnstiel (in prep) based on the dust evolution models from \cite{birnstiel2010}. This code solves the Smoluchowski coagulation equation \citep{smoluchowski1916} simultaneously with the transport of the grains. Dust particles collide with each other and are transported in the disk by different mechanisms:  thermal Brownian motion, vertical stirring and settling, turbulent mixing, and  azimuthal and radial drift. All of these velocities were taken into account to calculate the relative velocities before collision. The output after collision (coagulation, fragmentation, or erosion) is obtained by  comparing the relative velocities with the fragmentation speed $\vfrag$, a parameter that we set to 1\,m\,s$^{-1}$  in the entire disk based on recent laboratory experiments. 

Most of the transport and dynamics of the particles are regulated by the interaction of the dust particles with the gas. The dimensionless Stokes number quantifies the importance of the drag forces on the dynamics of the particles by comparing the stopping time of the particles to the Keplerian frequency at a given disk location. In the Epstein drag regime \citep{epstein1924} and near to the disk midplane, the Stokes number is defined as \citep{brauer2008, birnstiel2010}

\begin{equation}
\textrm{St}=  \frac{a\rho_s}{\sigmag}\frac{\pi}{2},
\label{eq:stokes}
\end{equation}

\noindent where $a$ is the grain size, $\rho_s$ is the intrinsic volume density of the particles, which we set to 1.2\,g\,cm$^{-3}$, and $\sigmag$ is the gas surface density. 

In our simulations we have the following general setup. We follow the dust evolution from 1000 years up to 1\,Myr. We assume a disk around a solar-mass star with an initial disk mass of 0.05\,$M_\odot$. Our radial grid is from 1 to 300\,au  with 300 cells that are logarithmically spaced. Initially, the gas-to-dust mass ratio is 100 and all the grains are one micron-sized particles. The particle size grid covers up to 2 meters with 180 cells, also logarithmically spaced. The temperature profile is a power law such that $T=160\,\mathrm{K}(r/\mathrm{au})^{-0.5}$, but we assume a floor value of 10\,K, implying that from $\sim$250\,au the temperature becomes constant. We assume an exponentially tapered power-law function for the initial gas surface density \citep{lynden1974}

\begin{equation}
        \Sigma_{\mathrm{g}}(r,t_0)=\Sigma_0\left(\frac{r}{R_c}\right)^{-\gamma} \exp\left[-\left(\frac{r}{R_c}\right)^{2-\gamma}\right]
  \label{eg:Sigma_disk}
\end{equation}

\noindent with $\gamma=1$ and a characteristic radius of $R_c=80$\,au. The gas surface density is viscously evolving following the diffusion equation obtained from the continuity equation and angular momentum conservation of the gas in protoplanetary disks and assuming Keplerian frequency everywhere in the disk \citep{pringle1981}, that is,

\begin{equation}
\frac{\partial \sigmag (r,t)}{\partial t}=\frac{3}{r}\frac{\partial}{\partial r}\left[ r^{1/2} \frac{\partial}{\partial r} (\nu \sigmag r^{1/2})\right], 
\label{eq:gas_evo}
\end{equation}

\noindent where $\nu$ is the disk kinematic viscosity. We use the parameterization by \cite{shakura1973}, such that

\begin{equation}
\nu=\alpha c_s  h \quad \textrm{with} \quad h=\frac{c_s}{\Omega},
\label{eq:viscosity}
\end{equation}

\noindent where $c_s$ is the isothermal sound speed, $h$ is the gas pressure scale height, and $\Omega$ is the Keplerian frequency. The parameter $\alpha$ is assumed to take values that range from $10^{-5}$ to  $10^{-2}$, the former being a value expected in the dead zones of protoplanetary disks \citep{nelson2013, klahr2014, lyra2014, flock2015, flock2020}, and the latter being values expected  in the magnetorotational instability active regions   \citep[e.g.,][]{fromang2006, davis2010, simon2011, flock2012, turner2014}. 

\subsubsection{Distinguishing the effects of gas diffusion versus dust (radial and vertical) diffusion}

Current dust evolution models of protoplanetary disks assume a dependence of turbulent velocities, vertical stirring, and radial diffusion of dust particles on the parameter $\alpha$ that regulates the gas viscous evolution. In our simulations, we assume non-isotropic turbulence and thus the parameters that control these physical mechanism are independent of each other. We use the symbols $\TurbPara$, $\delta_z$, and $\delta_r$, for the turbulence, settling, and radial diffusion, respectively. We always assume that particles cannot diffuse faster (radially or vertically) than the gas (i.e., $\delta_{r,z,\mathrm{t}}\leq\alpha$).

\paragraph{Vertical stirring and settling.} The velocities of very small particles ($\lesssim \SI{1}{\um}$ sized grains) are usually dominated by Brownian motion, but this becomes negligible as soon as particles grow to larger sizes \citep[e.g.,][see their Fig.~1]{birnstiel2016}. From micron- to sub-millimeter-sized dust grains, settling can be an import source for the velocity of the dust and their collisions. When the stirring by gas turbulence is neglected, the vertical velocity of grains that settle to the midplane ($v_{z, \mathrm{settling}}$) is given by balancing the gas drag force with the vertical gravitational force, $F_{\mathrm{grav}}=-m\Omega^2z$. This gives the terminal velocity for particles coupled to the gas ($\St<1$) $v_{z, \mathrm{settling}}=-z\Omega\St$ \citep{safronov1969, dullemond2004}.  
This implies that the  settling velocity decreases close to the midplane where the gas density increases, and it also increases for particles with higher Stokes number (for $\St<1$). Nonetheless, particles are also expected to be stirred up above the midplane by vertical turbulence. Vertical and radial turbulence may differ due to different instabilities, such as the Goldreich-Schubert Fricke instability \citep{nelson2013, stoll2014}. 

The relative velocities between particles due to settling depend on their height above the midplane, and we assume a Gaussian vertical mass distribution for which the scale height depends on the vertical stirring parameter $\delta_{z}$. The particle scale height $h_{\mathrm{d}}$ is given by \citep{youdin2007, birnstiel2010}

\begin{equation}
        h_{\mathrm{d}}(\St)=h \times \mathrm{min} \left( 1,\sqrt{\frac{\delta_z}{\mathrm{min}(\St,1/2)(1+\St^2)}}\,\right).
        \label{eq:dust_scaleheight}
\end{equation}

As demonstrated below, when the relative velocities due to turbulent motion and/or radial drift are low, relative velocities due to settling can dominate, in which case we use the settling limit obtained in Appendix~\ref{sec:GrowthBarriers}. 

Although we do not include the vertical evolution of the particle density distribution, we calculate the total volume dust density as a function of $h_d$ and use these 2D ($r$,$z$) particle density distributions in the radiative transfer models. The volume dust density is assumed to follow

\begin{equation}
        \rho_{\mathrm{d}}(r,\varphi,z, \St) = \frac{\Sigma_{\mathrm{d}}(r, \St)}{\sqrt{2\,\pi}\,h_{\mathrm{d}} (r, \St)}\,\exp \left( -\frac{z^2}{2\,h_{\mathrm{d}}^2(r, \St)} \right)\,,
        \label{eq:volume_density}
\end{equation}

\noindent where  $z = r\,\cos(\theta)$, with $\theta$ being a polar angle; and $\sigmad$ is the dust surface density.

\paragraph{Turbulent velocities.} For particles larger than submillimeter sizes, turbulence and radial drift can dominate the relative velocities, and both of them depend on the Stokes number. We approximate the turbulent velocities by \citep{ormel2007} 

\begin{equation}
\vfrag^2\sim3\TurbPara\St~c_s^2.
\label{eq:v_turb}
\end{equation}

When we assume that turbulence dominates the relative velocities of particles and that sticking collisions occur below the fragmentation speed $\vfrag$, the maximum grain size $a_{\mathrm{f, turb}}$ is given by the so-called fragmentation barrier \citep{birnstiel2010,birnstiel2012},

\begin{equation}
        a_{\mathrm{f, turb}}=\frac{2}{3\pi}\frac{\sigmag}{\rho_s \TurbPara}\frac{\vfrag^2}{c_s^2}.
  \label{eq:afrag}
\end{equation}

We assume values of  $\delta_t$ that are either equal to or lower than $\alpha$. Physically, this would imply that what sets the gas viscous evolution may be detached from what dominates the turbulent motion of the grains. This is possible for instance in the weakly ionized regions (or dead zones), where  the magneto-rotational instability (MRI) is not effective as expected in the dense midplane regions, while what drives the gas accretion and the angular momentum transport in the disk may be driven by other and more effective physical mechanism such as a MHD disk wind \citep[e.g.,][]{bai2016}.

\begin{table*}
\begin{center}
\caption{Summary of the models and results without pressure bumps (Fig.~\ref{fig:no_bumps}) Columns 2, 3, 4, and 5 show the values taken for the parameters that regulate the gas viscous evolution ($\alpha$), radial diffusion ($\delta_r$), turbulent mixing ($\delta_t$), and vertical settling/stirring ($\delta_{z}$), respectively. Column 6 shows a tick~mark when pebbles ($a>$1\,mm) have formed in most of the disk, or a cross  otherwise. Column 7 summarizes the relative velocity that sets the maximum grain size, which is always in the fragmentation regime at 1\,Myr of evolution. Columns 8 and 9 show the synthetic millimeter fluxes at 1.3\,mm and 3.0\,mm, respectively. Column 10 is the spectral index. Columns 11 and 12 show the dust radius that encloses 68\% or 90\% of the 1.3\,mm flux.} \label{tab:summary_table}
\begin{tabular}{ |l|c|c|c|c|c|c|c|c|c|c|c|} 
    \hline
    \hline
    (1)&(2)&(3)&(4)&(5)&(6)&(7)&(8)&(9)&(10)&(11)&(12)\\
    \small{Model} & $\alpha$ &$\delta_r$ &   $\TurbPara$ & $\delta_{z}$ &  \tiny{Pebble} & $a_{\mathrm{max}}$& $F_{1.3\mathrm{mm}}$ & $F_{3.0\mathrm{mm}}$ & $\alpha_{\mathrm{mm}}$ & $R_{\mathrm{dust, 68\%}}$ & $R_{\mathrm{dust, 90\%}}$\\
          &&&&&\tiny{formation?}& \small{set by:} & [mJy] & [mJy] & & [au]& [au]  \\
          &&&&&\tiny{($a>$1\,mm)}&&&&&&  \\
    \hline
    1     & $10^{-2}$ &           $10^{-2}$  &    $10^{-2}$           & $10^{-2}$ &     \xmark     &\small{turbulence} &127.2 &6.0 &3.7 &120 &257\\
    \hline
    2     & $10^{-5}$ &           $10^{-5}$  &    $10^{-5}$           & $10^{-5}$ &     \cmark     &\small{drift}&295.2 &19.3 &3.3 &57 &91\\
    \hline
    3     & $10^{-2}$ &           $10^{-2}$  &    $10^{-4}$           & $10^{-3}$ &     \cmark     &\small{settling/drift}&198.4 &10.3 &3.5 &41 &82\\
    \hline
    4     & $10^{-2}$ &           $10^{-2}$  &    $10^{-5}$           & $10^{-2}$ &     \xmark     &\small{settling}&163.4 &7.3 &3.7 &55 &198\\
    \hline
    5     & $10^{-2}$ &           $10^{-3}$  &    $10^{-5}$           & $10^{-2}$ &     \xmark     &\small{settling}&150.4 &7.0 &3.7 &37 &95\\
    \hline
    6     & $10^{-3}$ &           $10^{-3}$  &    $10^{-5}$           & $10^{-3}$ &     \cmark     &\small{settling/drift}&295.5 &21.2 &3.2 &58 &87\\
    \hline
    7     & $10^{-3}$ &           $10^{-3}$  &    $10^{-3}$           & $10^{-3}$ &     \xmark     &\small{turbulence}&313.6 &17.0 &3.5 &47 &91\\
    \hline
    8    & $10^{-3}$ &           $10^{-4}$  &    $10^{-5}$            & $10^{-4}$ &     \cmark     &\small{drift}&304.2 &19.2 &3.3 &65 &96\\
    \hline
    9    & $10^{-4}$ &           $10^{-4}$  &    $10^{-4}$          & $10^{-4}$ &       \cmark     &\small{turbulence/drift}&327.9 &22.3 &3.2 &61 &95\\
    \hline
    \hline
\end{tabular}
\end{center}
\end{table*}

Growth can also be limited by radial drift because the timescale of radial drift is short. When the growth timescales are balanced with the drift timescales \citep{klahr2006} and when relative particle speeds are assumed to be dominated by turbulent motion, the drift barrier is given by \citep{birnstiel2012}

\begin{equation}
        a_{\mathrm{drift}}=\frac{2 \sigmad}{\pi\rho_s}\frac{v_K^2}{c_s^2}\left \vert \frac{d \ln P}{d\ln r} \right \vert^{-1},\end{equation}

\noindent where $P$ is the disk gas pressure, and $v_K$ is the Keplerian speed. 

The two-population model presented by \cite{birnstiel2012} takes advantage of top-heavy size distributions and of the increase of the radial drift speed with size (for Stokes numbers below unity). 
Under these conditions, particle evolution can be simplified by assuming that a particle flux dominates large species and a small species that is radially transported mostly by gas motion and particle diffusion.  \cite{birnstiel2012} assumed in their model that fragmentation occurs as a result of relative turbulent motion or relative radial drift. For the drift limit, it is assumed that turbulent motion dominates. 
However, for low values of $\TurbPara$, that is, lower relative turbulent velocities, and/or low particle fragmentation speeds $\vfrag$, the dominating source of relative velocity can switch to relative vertical settling or relative radial drift in the fragmentation and drift regime. This means that in each regime turbulence, settling, or radial drift can also dominate, and thus each regime can have up to three different barriers, as explained in Appendix~\ref{sec:GrowthBarriers}. These limits can be used to further generalize the two-population model presented in \cite{birnstiel2012}, which allows us more freedom in choosing the viscosity parameter of the gas $\alpha$, the particle turbulence parameter $\TurbPara$, and the vertical ($\DiffZ$) and radial ($\delta_r$) particle diffusion parameters separately.

\paragraph{Radial diffusion.} For the dust transport, the advection-diffusion equation of the dust surface density ($\sigmad$) is solved for each grain size, given by \citep[e.g.][]{birnstiel2010}

\begin{equation}
        \frac{\partial \sigmad}{\partial t} + \frac{1}{r}\frac{\partial}{\partial r}\left( r \sigmad v_{\mathrm{r,d}}\right)-\frac{1}{r}\frac{\partial}{\partial r} \left[r \sigmag D_\mathrm{d} \frac{\partial }{\partial r}\left(\frac{\sigmad}{\sigmag}\right)\right]=0,
  \label{eq:dustevo}
\end{equation}

\noindent where $D_\mathrm{d}$ is the dust radial diffusivity, which is assumed to be \citep{youdin2007}

\begin{equation}
        D_\mathrm{d}=\frac{\delta_r\,c_s\,h}{1+\St^2}.
        \label{eq:diffusion}
\end{equation}

In the classical models, $\delta_r=\alpha$, but here we assumed independent values with the same motivation as for  $\delta_t$ or $\delta_{z}$, that is, the radial diffusion of the grains is independent of the physical mechanism that causes the viscous evolution of the gas.

\subsubsection{Combination of the $\TurbPara$, $\delta_{z}$, $\delta_r$, and $\alpha$ parameters}

Table~\ref{tab:summary_table} summarizes the combination of $\delta_t$, $\delta_{z}$, $\delta_r$, and $\alpha$ that we assumed for the models of this paper. Models 1 and 2 were the two extremes in our set of simulations. Model 1 considered a disk in which the parameter for the viscous evolution of the gas was set to the maximum value ($10^{-2}$) obtained in the active regions of disks from MRI calculations \citep[e.g.,][]{turner2014}, and in which the remaining analyzed parameters had the same values. On the other hand, model 2 was the other extreme, in which the parameter for the viscous evolution of the gas and all the others were set to a very low value that is expected in a disk in wich MRI is inefficient ($10^{-5}$). 
The values for the parameters to explore in models 3 to 9 were such that they covered a transition from the two extremes of model 1 and 2. More specifically, models 7 and 9 also assumed that all the four parameters are equal, and the values were $10^{-3}$ and $10^{-4}$, respectively. The remaining models (models 3, 4, 5, 6, and 8) mixed different values to resemble different radial/vertical diffusion and/or turbulent velocities of the particles.

\begin{figure*}
 \centering
        \includegraphics[width=18cm]{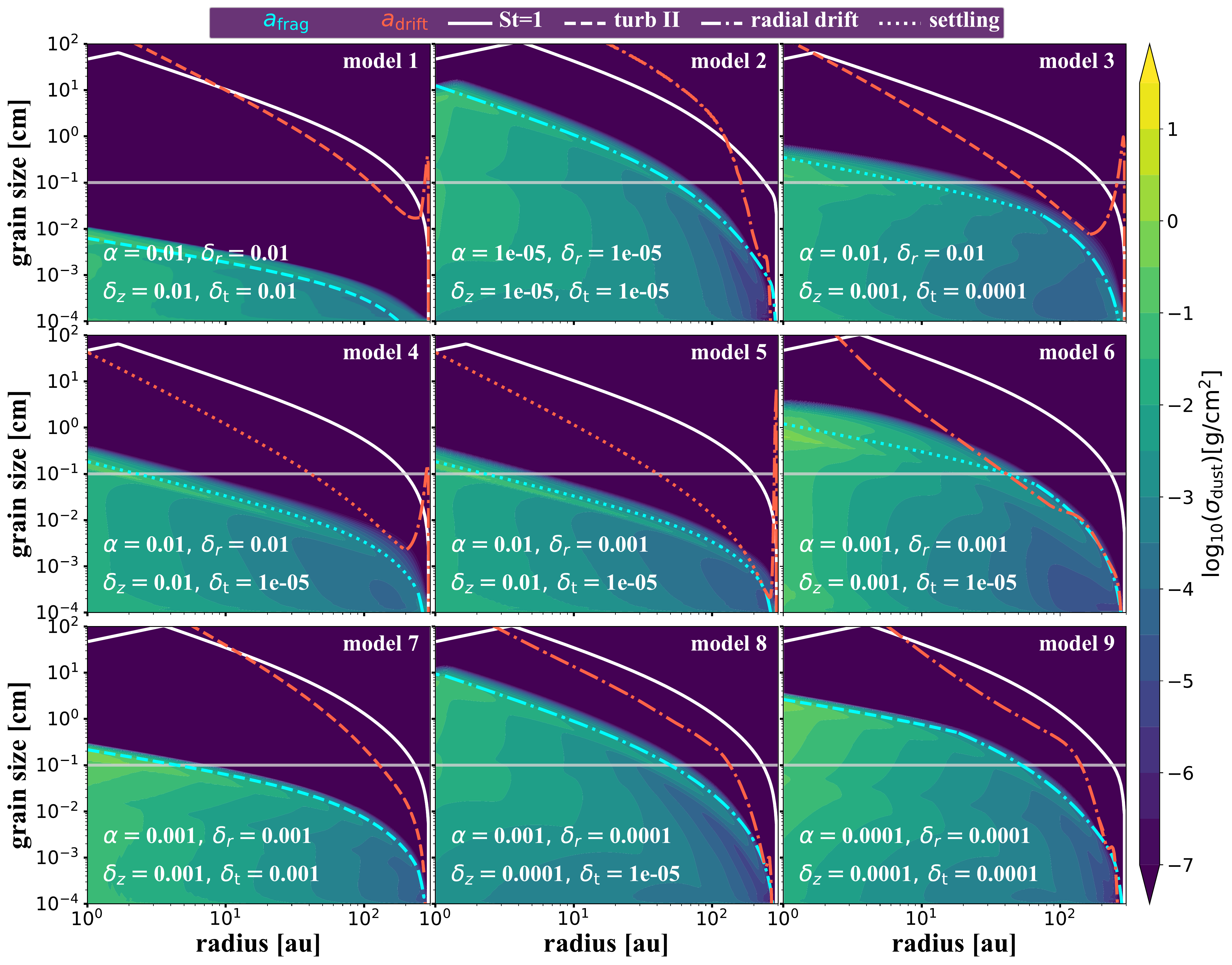}\\  \vspace{-0.2cm}
   \caption{Particle density distribution after 1\,Myr of evolution as a function of distance from the star and grain size of the models without any pressure bump (Table~\ref{tab:summary_table}). The solid white line represents a Stokes number of unity. Cyan and red show the fragmentation (Eq.~\eqref{eq:fragLim}) and drift (Eq.~\eqref{eq:driftLim}) limit. The style of the line shows what dominates the relative velocity: turbulence (Eqns.~\eqref{eq:StFragTurb} and \eqref{eq:StDriftTurb2}), radial drift (Eqns.~\eqref{eq:StFragR}, and \eqref{eq:Stdriftr}), and settling (Eqns.~\eqref{eq:StFragZ} and \eqref{eq:StDriftZ}). We do not include the turbulence I regime or the azimuthal drift because they do not dominate the relative velocities in any case for the shown snapshot. The horizontal gray line shows $a=1\,$mm, and we call pebbles all particles that are larger than this. }
   \label{fig:no_bumps}
\end{figure*}

\subsubsection{Inclusion of a pressure bump}

We followed the prescription introduced by \cite{dullemond2018}, which was also used by \cite{stammler2019}, to include a pressure bump in the disk. This assumes a bump that originates from variations of the  viscosity parameter ($\alpha$), modeled as

\begin{equation}
  \alpha\prime\left(r\right) = \frac{\alpha}{F\left( r \right)},
  \label{eq:bump0}
\end{equation}

\noindent where  $F\left( r \right)$ is given by

\begin{equation}
  F \left( r \right) = \exp \left[ -f \exp \left( -\frac{\left( r - r_\mathrm{p} \right)^2}{2w_\mathrm{gap}^2} \right) \right].
  \label{eq:bump1}
\end{equation}

From previous inspections of the results of models without bumps, we performed six simulations with this pressure bump, as summarized in Table~\ref{tab:summary_table2}. In these models, only $\alpha$ varied with radius, and the remaining parameters were set as before for comparison (see Table~\ref{tab:summary_table2} for more details). We assumed the  position of the bump at $r_{\mathrm{p}}=40$\,au, with an amplitude of $f=2$ \citep[as in][]{dullemond2018, stammler2019}, and a width of $w_{\mathrm{gap}}=5\,$au, which is approximately twice the local pressure scale height at $r_{\mathrm{p}}=40$\,au. However, the pressure bump itself formed farther out, at $\sim52\,$au, where the pressure scale height is $\sim$3.5\,au.

\subsection{Radiative transfer calculations}

To compute synthetic images and spectral energy distributions (SEDs) for each model, we considered the resulting vertical particle  distribution (Eq.~\eqref{eq:volume_density}) after 1\,Myr of evolution as input to the radiative transfer calculations, for which we used \textsc{RADMC-3D} \citep{dullemond2012}. We calculated the opacity of each grain size considering porous spheres with a dust mixture composed of astronomical silicates \citep{draine2003}, carbonaceous material \citep{zubko1996}, and water ice \citet{warren2008}, as in \citet{ricci2010}. We assumed a blackbody radiation field from the central star as the radiation source and used $1\times10^{7}$ photons for our calculations.

We obtained the temperature profile for each grain size and calculated scattered-light images at $\SI{1.25}{\um}$ and 1.3\,mm. We assumed a disk inclination of 20$^\circ$, a position angle of zero, and a distance to the source of 140\,pc as the typical distances to nearby star-forming regions, such as Lupus, Taurus, and Ophiuchus. The synthetic scattered-light images were computed including the full treatment of anisotropic scattering with polarization, as in \cite{kataoka2015} and \cite{pohl2016}. Afterward, the images were convolved by a Gaussian point-spread function (with a size of 0.04'' $\times$ 0.04'') at the two wavelengths, which is similar to the angular resolution that can currently be obtained with Spectro-Polarimetric High-contrast Exoplanet REsearch at the Very Large Telescope (SPHERE/VLT) and ALMA. The intensity profiles created at $\SI{1.25}{\um}$ were then multiplied with the square of their distance to the star to compensate for the stellar illumination. From the intensity profiles at 1.3\,mm we calculated the radius that enclosed either 68\% or 90\% of the emission by assuming that the observation sensitivity provides a signal-to-noise ratio with respect to the maximum of the emission (peak) of 1000. 

\begin{figure*}
 \centering
        \includegraphics[width=18cm]{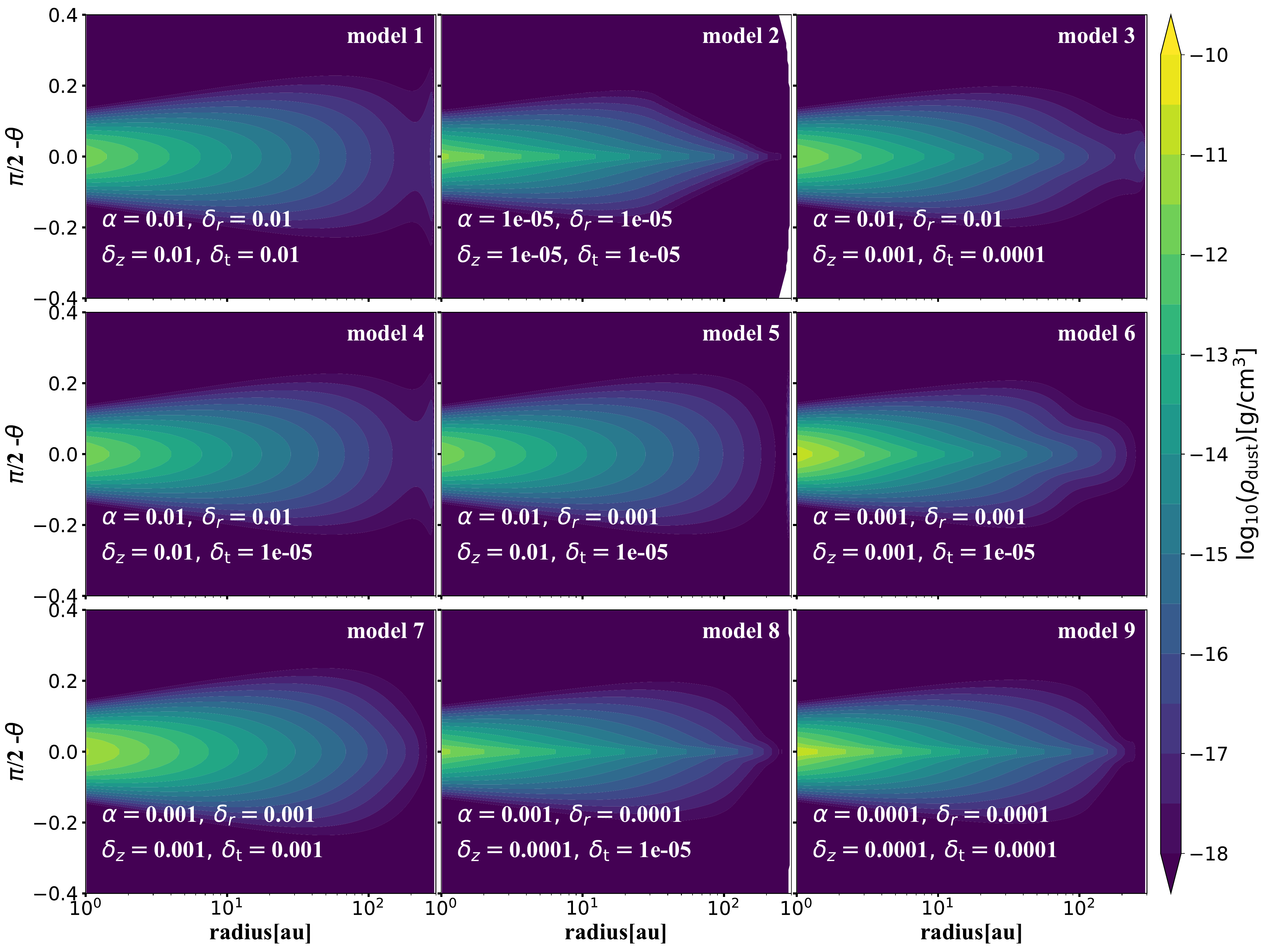}\\  \vspace{-0.2cm}
   \caption{Vertical and radial particle density distribution for all grain sizes of models without any bumps after 1\,Myr of evolution.}
   \label{fig:no_bumps_3D}
\end{figure*}

\section{Results} \label{sect:results}
\subsection{Models without pressure bumps}
The particle density distribution as a function of distance from the star and grain size of the models without any pressure bump (Table~\ref{tab:summary_table}) is shown in Fig.~\ref{fig:no_bumps}. All panels show the results at 1\,Myr of evolution (as the typical age of class II protoplanetary disks in nearby star forming regions like Taurus and Lupus) and in all cases the fragmentation speed is $\vfrag=1\mathrm{m\,s}^{-1}$.

Models\,1 and \,2 are the two extremes of our assumptions. For all the parameters that set the gas and dust diffusion as well as relative turbulent speeds, model 1 adopted the same value of $10^{-2}$ , and model\,2 corresponds to the case of $10^{-5}$. 
As expected in model\,1, pebble ($a>1\,$mm) formation did not occur. 
The maximum grain size in the case of model 1 was set by high velocities because turbulence induced the fragmentation barrier ($a_{\mathrm{f, turb}}$, Eq.~\ref{eq:afrag}), and the maximum grain size was approximately 0.01\,cm at 1\,au and about $\SI{10}{\um}$ in the outer part of the disk (at about 100\,au). On the other hand, model\,2 allowed the formation of pebbles in most of the disk, where the maximum grain size was set by fragmentation, for which the radial drift velocities dominate. 

The results of model 1 and 2 are expected because $a_{\mathrm{f, turb}}$ depends quadratically on the fragmentation velocity. When we assume a fragmentation velocity that is one order of magnitude lower than previously used, theory requires us to decrease the dust turbulence by at least two orders of magnitude if turbulence is the main source of relative velocities.  
Model\,9 was the same as model\,2, but with $10^{-4}$. It shows that the maximum grain size for $r<50\,$au is set by fragmentation by turbulence and by drift in the outer regions. In this case there is still some pebble formation. 
Model 7 assumed all gas and dust diffusion parameters to be equal to $10^{-3}$ , and pebble formation did not occur in most of the disk, as in model\,1. The first conclusion of these numerical experiments therefore is that with isotropic turbulence and $\vfrag=1\mathrm{m\,s}^{-1}$, pebble formation can occur in disks when the gas and dust diffusion are low, with values of $10^{-5}-10^{-4}$. 

However, the main motivation for the other experiments of this work was first that when $\alpha$ has values of  $10^{-5}-10^{-4}$ , the disk accretion rate at million-year timescales are expected to be $10^{-12}-10^{-11}\,M_\odot\,\mathrm{yr}^{-1}$, which is much lower than the typical accretion rates in different star-forming regions of different age \citep[1-10\,Myr, see, e.g.,][]{manara2020}. For values of $\alpha$  of  $10^{-3}-10^{-2}$, the accretion rates values agree with the mean values from observations (i.e., $10^{-9}-10^{-10}\,M_\odot\,\mathrm{yr}^{-1}$). In the remaining numerical experiments we therefore kept $\alpha=10^{-2}$ or  $\alpha=10^{-3}$ and varied the parameters responsible for the radial and vertical particle diffusion as well as the parameter that controls the collision speed due to turbulent motion. 

\begin{figure*}
 \centering
        \includegraphics[width=18cm]{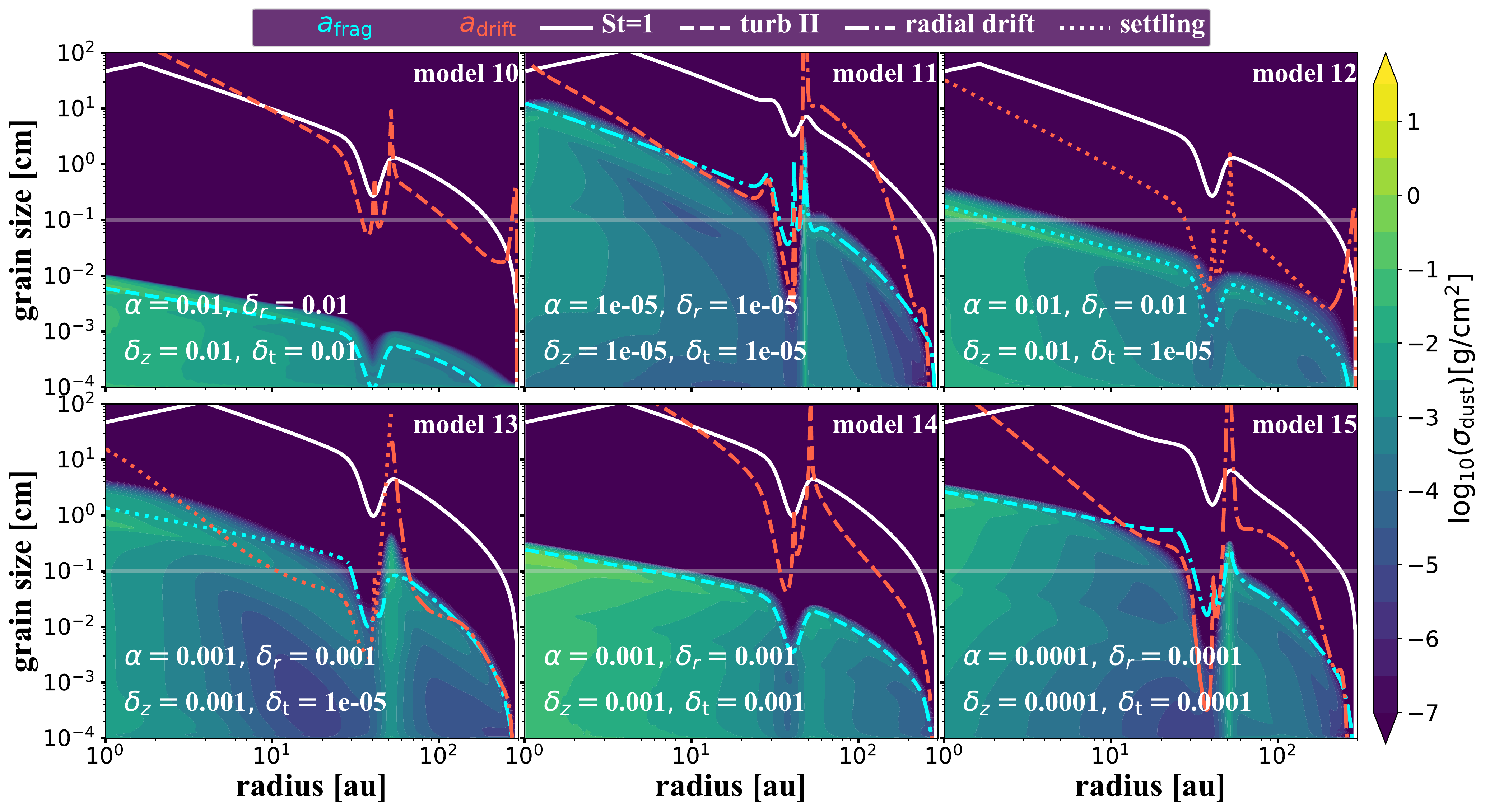}\\  \vspace{-0.2cm}
   \caption{Particle density distribution of the models with a pressure bump in the outer part of the disk (dip located at 40\,au) after 1\,Myr of evolution. We follow the same procedure as in Fig.~\ref{fig:no_bumps} for the barriers.}
   \label{fig:bumps_2D} 
\end{figure*}

The second motivation for these numerical experiments was that several protoplanetary disks appear to be flared when observed in scattered light \citep{avenhaus2018}. Assuming that we trace micron-sized particles with these observations, and taking values of $\delta_{z}=10^{-5}-10^{-4}$, micron-sized particles would be mainly settled in the midplane, contradicting observations. 

The third motivation is that the width of some rings in protoplanetary disks observed at very high resolution with ALMA have been resolved \citep[e.g.,][with values of 4-20\,au]{dullemond2018}. This means that there must be radial diffusion inside pressure bumps to widen the distribution of millimeter-sized particles. All the observed rings would otherwise be “infinitely” narrow (dust grains would move to the location of the pressure maximum).

Our models 3, 4, 5, and 6  maintained high values of $\alpha$, $\delta_r$, and $\delta_{z}$ to guarantee accretion-rate values, width of rings, and dust aspect ratios that are in accordance with observations, while the parameter that controls the turbulent velocities ($\delta_t$) of the particles was low. Model\,8 is an experiment in which we also decreased $\delta_{z}$ to $10^{-4}$ because by previous inspections of the models, we realized that settling can set the maximum grain size and limit pebble formation. Table~\ref{tab:summary_table} shows when pebble formation occurred in our models and what set the maximum grain size in each case. At 1\,Myr of evolution, all the cases are in the fragmentation regime, but the relative velocity that dominates and sets the maximum grain size is different.

From our numerical experiments in which we used different values for $\TurbPara$, $\delta_{z}$, $\delta_r$, and $\alpha$ (i.e., models 3, 4, 5, 6, and 8), the maximum grain size was set either by settling or radial drift. In models 4 and 5, pebble formation is not possible in almost the entire disk ($r>2\,$au) when $\delta_{z}=10^{-2}$. The reason is the average settling velocities dominate the total relative velocities and set the maximum grain size. Models 4 and 5 are identical except for the radial diffusion, which is lower in model 5. This creates a small effect in the particle distribution, but the radial extension of the large grains is slightly larger for model 4, as we discuss in the Sect.~\ref{sect:discussion} when we compare the results with observations.

For models 3 and 6 with $\delta_z=10^{-3}$, the maximum grain size was mostly set by settling and was set by drift in only the farthest regions. In these two cases, pebble formation occurred for $r<10\,$au in model 3 and for $r<50\,$au for model 6. Model\,6 produced pebbles in a wide range of the disk. This model is a good compromise for a comparison with observations, that is, an $\alpha=10^{-3}$ that gives accretion rates of $~10^{-9}\,M_\odot\,\mathrm{yr}^{-1}$,  a $\delta_{z}=10^{-3}$ that can lead to enough vertical stirring to produce a high dust-aspect ratio similar to those observed with SPHERE for T-Tauri and Herbig disks, and a $\delta_r$ that can lead to radial diffusion and a wider distribution of pebbles (also for the potential presence of a pressure bump).

No direct observational constraint is available that could help us to fix $\TurbPara$. Indirect measurements of the disk turbulence have been made using ALMA observations by searching for variations from Keplerian motion of the gas emission that may come from turbulence. \cite{flaherty2015, flaherty2017, flaherty2020, teague2016, teague2018} found low turbulence values for four disks ($\lesssim 10^{-3}$).  \cite{flaherty2020} showed that DM\,Tau is the only system observed so far with high turbulence.

\begin{figure*}
 \centering
        \includegraphics[width=18cm]{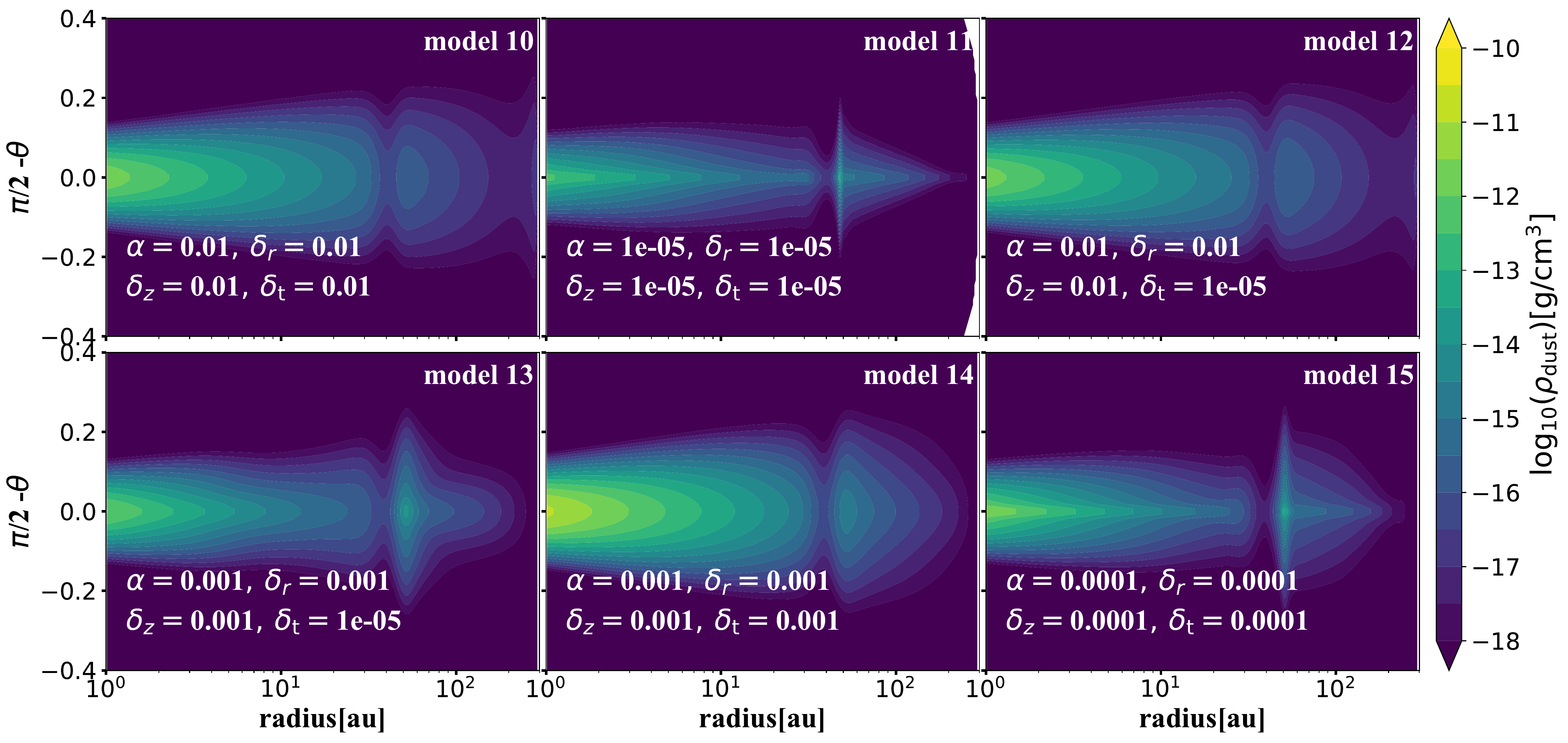}\\  \vspace{-0.2cm}
   \caption{Vertical and radial particle density distribution of all grain sizes of the models in which a pressure bump is located in the outer disk.}
   \label{fig:bumps_3D}
\end{figure*}

\begin{table*}
\begin{center}
\caption{Summary of the models and results with a pressure bump (Fig.~\ref{fig:bumps_2D}). In Column 3 a tick~mark is given when efficient trapping occurs in the pressure bump. A cross is shown otherwise. \label{tab:summary_table2}}
\begin{tabular}{ |l|c|c|c|c|c|c|c|}
    \hline
    \hline
    (1)&(2)&(3)&(4)&(5)&(6)&(7)&(8)\\
    Model & Description          & Trapping? & $F_{1.3\mathrm{mm}}$ & $F_{3.0\mathrm{mm}}$ & $\alpha_{\mathrm{mm}}$ & $R_{\mathrm{dust, 68\%}}$ & $R_{\mathrm{dust, 90\%}}$ \\
          &                      &           & [mJy] & [mJy] & & [au]& [au]  \\
    \hline
    10    & as model 1 with bump & \xmark &119.4 &5.7 &3.6 &141 &263\\
    \hline
    11    & as model 2 with bump & \cmark &203.8 &12.4 &3.3 &75 & 100\\
    \hline
    12    & as model 4 with bump & \xmark &148.8 &6.8 &3.7 &78 &213\\
    \hline
    13    & as model 6 with bump & \cmark &331.2 &24.1 &3.1 &65 & 88\\
    \hline
    14    & as model 7 with bump & \xmark &286.6 &16.8 &3.4 &61 &97\\
    \hline
    15    & as model 9 with bump & \cmark &247.7 &16.0 &3.3 &73 &101\\
    \hline
    \hline
\end{tabular}
\end{center}
\end{table*}

Figure~\ref{fig:no_bumps_3D} shows the 2D (radial and vertical) particle density distribution assumed for the radiative transfer calculations and obtained using Eq.~\eqref{eq:volume_density}. We show the vertical and radial particle density distribution considering all the grain sizes. 
The greatest difference is seen between our two extremes model 1 ($\alpha, \delta_{r,z,t}=10^{-2}$) and model 2 ($\alpha, \delta_{r,z,t}=10^{-5}$), where model 2 clearly presents much less stirring of dust particles from the midplane, making the dusty disk thinner. 
Another clear result from Fig.~\ref{fig:no_bumps_3D} is that when the maximum grain size is set by  the fragmentation limit due to radial drift, the disk becomes thinner in all grains. This occurs in all the disk for models 2 and 8 in which radial drift dominates relative velocities everywhere, and from the location where the barrier is set by radial drift and outward for models 3, 6, and 9. In addition, in models where $\delta_{z}\leq 10^{-4}$ (models 2, 3, 8, and 9),  the particle density distribution is much more dominated by the midplane than in the other models.

\subsection{Models with a pressure bump in the outer disk}
We chose our two extreme cases (model 1 and 2) for the inclusion of a pressure bump, together with the other two cases in which all the gas and dust diffusion parameters are constant (models 7 and 9). In addition to these cases, we included two other cases, one in which the maximum grain size was set by settling in the whole disk (model\,4). The other case was model\,6, in which pebble formation occurred, and the maximum grain size was set by a mix of settling and drift. The summary of the models that include a pressure bump as described in Eqns.~\eqref{eq:bump0} and ~\eqref{eq:bump1} is listed in Table~\ref{tab:summary_table2}.  

The particle density distribution as a function of distance from the star and grain size for the case of a gap at 40\,au and a pressure bump at the edge of that gap is shown in Fig~\ref{fig:bumps_2D}. Because this gap is formed in the $\alpha$ viscosity, any variation in the value of $\alpha$ changes the timescale within which this gap and bump are formed. Because all the plots are shown at 1Myr, in model 11 when $\alpha$ is the lowest, the gap is still opening and the amplitude of the bump is lowest. The gap is also shallower for model 15 ($\alpha=10^{-4}$) than in the rest of the models.

There is effective trapping, reflected by the significant particle accumulation at the pressure maximum, when pebble formation occurs (models 11, 13, and 15). In the other cases, the strong radial mixing (both through the radial particle diffusion parameter and the radial gas velocity by which small particles are dragged along) does not lead to significant trapping and hence to no significant accumulation of  particles at the pressure maximum (models 10, 12, and 14). 

When there is no efficient trapping (models 10, 12 and 14), there is a small decrease in the maximum particle size at the location of the gap because the fragmentation barrier is proportional to the gas surface density (regardless of whether it is set by turbulence or settling). When there is no trapping and the maximum grain size is set by settling (model 12), the decrease in particle size at the gap location is smaller. 

For effective particle trapping, model 11 ($\alpha$, $\delta_{r,z,\mathrm{t}}=10^{-5}$) shows a situation where grain growth is very efficient in the pressure maximum and particles grow to the maximum size that is allowed in our grid (2\,m) at that location. The reason is that the relative velocity that dominates fragmentation is radial drift, which is zero at pressure maximum. This has been seen before in simulations with particle traps and low turbulence and diffusion of the grains \citep[e.g.,][]{pinilla2015}. Models 13 ($\alpha$, $\delta_{r,z}=10^{-3}$, and $\TurbPara=10^{-5}$) and 15 ($\alpha$, $\delta_{r,z,\mathrm{t}}=10^{-4}$) where effective trapping occurs, settling or turbulence set the maximum grain size inside the pressure bump, respectively. The width of the dust accumulation inside the gap decreases for lower $\delta_r$, that is, model 11 ($\delta_r=10^{-5}$) shows the narrowest accumulation, followed by model 15  ($\delta_r=10^{-4}$), and then by model 13  ($\delta_r=10^{-3}$).

Figure~\ref{fig:bumps_3D} shows the vertical and radial distribution integrated over all the grain sizes for each model that includes a pressure bump. When there is no efficient dust trapping (models 10, 12, and 14), the decrease in particle density distribution is small, which is more evident in the upper layers of the disk, but it is still noticeable at the midplane. 

For effective trapping (models 11, 13, and 15), the depletion of grains extends from the surface layers to the midplane, especially in cases 11 and 15, where the radial diffusion and vertical stirring are low. At the location of the trap, small grains clearly increase as a result of continuous fragmentation of dust particles, as seen before in \cite{pinilla2012} and \cite{pinilla2015}, for example, but also in 2D dust evolution simulations by \cite{joanna2019}. Because of this local increase in small particles, their vertical distribution increases through vertical stirring. This is more clear in models 13 and 15, where the vertical stirring parameter is higher than in model 11. Model 13 has higher radial diffusion, which causes a significant number  of grains throughout the gap, more than in models 11 or 15. 

\section{Discussion} \label{sect:discussion}

\subsection{Millimeter fluxes and spectral indices}
Tables~\ref{tab:summary_table} and~\ref{tab:summary_table2} provide the expected millimeter fluxes at 1.3\,mm (ALMA Band\,6) and 3.0\,mm (ALMA Band\,3) after taking the particle density distribution after 1\,Myr of evolution. They then use radiative transfer simulations to create the SEDs. 

Model 1, in which the fragmentation barrier is the lowest and where pebble formation does not occur in the disk, has the lowest millimeter fluxes. In general, the cases where pebbles are not formed or they are formed in just a small region of the disk have the lower millimeter fluxes, except for model\,7, in which all the gas and dust diffusion parameters were set to $10^{-3}$. None of the models show a correlation between the velocities that set the maximum grain size and the obtained fluxes. All of the fluxes are within the observed values of the disks in different star-forming regions around T-Tauri and Herbig stars \citep[e.g.,][]{ricci2012, tripathi2017, andrews2018}. 

\begin{figure}
 \centering
        \includegraphics[width=9cm]{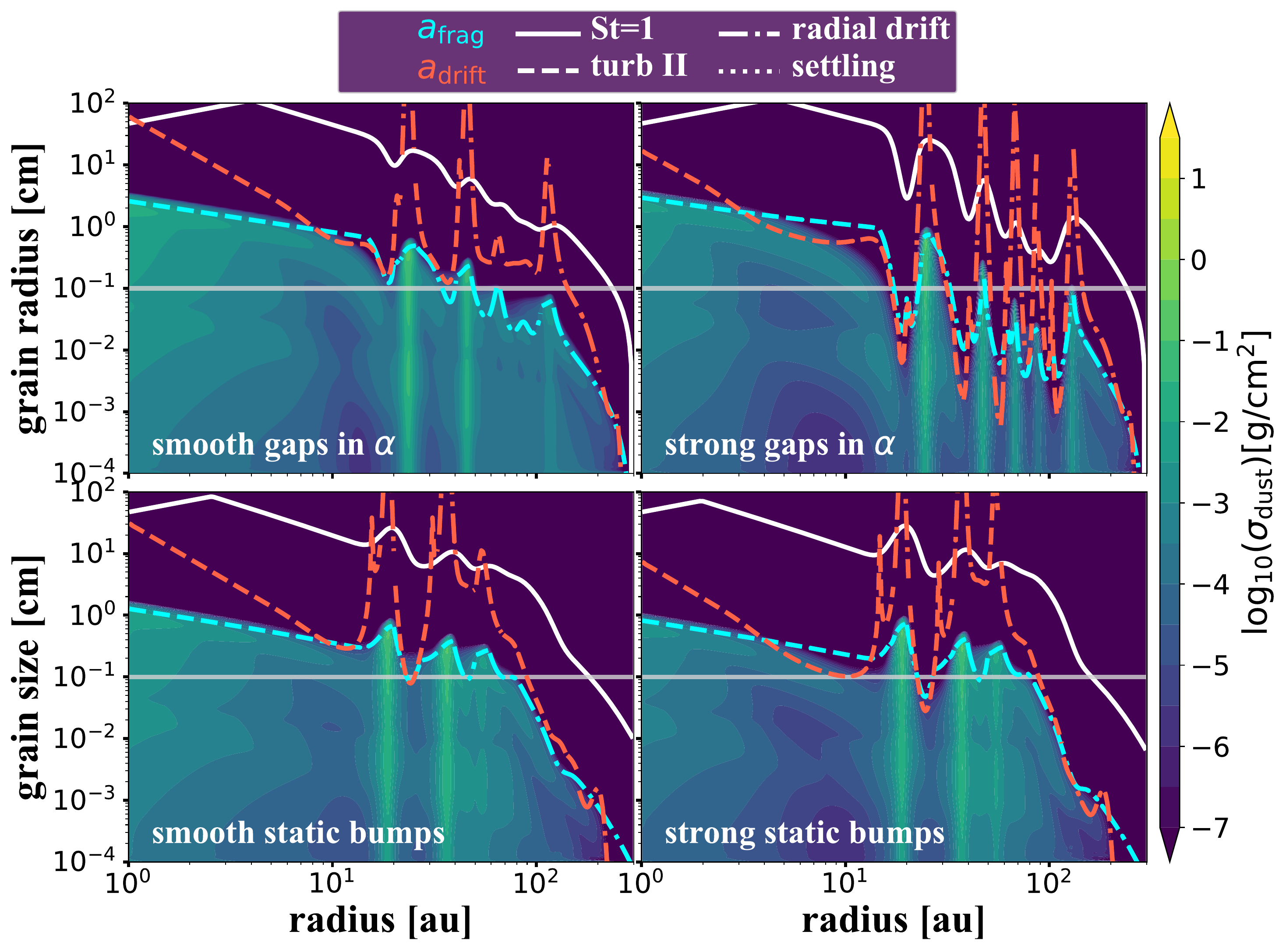}\\
   \caption{As Fig.~\ref{fig:no_bumps}, but with multiple pressure bumps induced either by variations in $\alpha$ and different gap depths (top panels), or introduced in the gas surface density that remains constant over time (bottom panels). The growth barriers follow the same convention as in Fig.~\ref{fig:no_bumps}. In these cases, $\alpha,\delta_{r,z,\mathrm{t}}=10^{-4}$ for all four simulations.}
   \label{fig:multiple_bumps}
\end{figure}
 
When we compare the models with and without a bump in the disk, in all of the cases (except for model 6 vs. model 13), the millimeter fluxes decrease when the bump is included. When trapping occurs, this originates from (i) grains growing to decimeter objects, which have very low opacities and contribute very little to the total fluxes (model 2 vs. model 11), or/and (ii) the emission inside the trap becoming optically thick (as in the case of model 15). In model 13 (model 6, but with a bump), grains in the trap have a maximum size of $\sim$1\,cm and the emission is optically thin such that the millimeter fluxes slightly increase with the trapping, as expected because fewer grains are drifting toward the star. In the cases without efficient trapping, the fluxes decrease compared to the same cases without a bump. In all these cases, we have a local decrease of the maximum grain size at the gap location, which helps to decrease the total millimeter flux.

For all the models, we obtain the millimeter spectral index ($\alpha_{\mathrm{mm}}$) taking the fluxes at 1.3\,mm and 3.0\,mm, such that $\alpha_{\mathrm{mm}}=\log_{10}(F_{1.3\mathrm{mm}}/F_{3.0\mathrm{mm}})/\log_{10}(3.0/1.3)$. All the spectral indices in our work are much higher than observed \citep[e.g.,][]{ricci2012, pinilla2017b}. As shown in \cite{pinilla2012}, high spectral indices from dust evolution models are expected when radial drift is included. Even when pebble formation occurs and the millimeter fluxes agree with observations, the number of pebbles is not enough to have low spectral indices. Even for models with effective trapping, the values of the spectral index remain similar as in models without a bump, implying that this bump does not maintain enough pebbles to decrease the values of the spectral index (only 1-2\% of the total dust mass is in pebbles). To solve the discrepancy, strong multiple bumps can assist in decreasing $\alpha_{\mathrm{mm}}$ \citep{pinilla2012}. Alternatively, when only one single bump is included, it has to be strong enough, as the one expected from a giant planet, to decrease the spectral index \citep{pinilla2014}. In this case, the spectral index depends on where the bump is located, becoming larger when the bump is located farther out. Another factor that can affect the final spectral indices is the lifetime of the pressure bumps because in previous works \citep[e.g.,][]{pinilla2012,pinilla2013, pinilla2014}, pressure bumps were included from the beginning of the simulation and they are static, while in the present work the gap and bump grow on viscous timescales. Both completely static bumps and bumps that slowly grow with time have their limitations. A prescription of a bump in $\alpha$ allows us to have mass accretion onto the star as well as viscous spreading, but for very low values of $\alpha$ it can take more than 1\,Myr to reach the final amplitude of the bump, which may not be the case of planets that cause the bump formation, for instance. A more realistic approach would include the viscous evolution and the grow, appearance, and disappearance of a number of bumps with a given frequency during the disk lifetime.

To test whether multiple gaps as assumed in this paper versus static multiple bumps affect the spectral index, we performed four more models considering the $\alpha$ and $\delta_{r,z,\mathrm{t}}$ from model 9 (i.e., $\alpha, \delta_{r,z,\mathrm{t}}=10^{-4}$), and we took five gaps or bumps with the same width of 2 local pressure scale heights. They were located at 20, 40, 60, 80, and 100\,au. We list these models below.

\begin{itemize}
    \item A model with five gaps as prescribed in this paper (Eq.~\ref{eq:bump0} and Eq.~\ref{eq:bump1}) with $f=2$ (name: smooth gaps in $\alpha$).
    \item As the previous model, but with $f=4$ (name: strong gaps in $\alpha$).
    \item A model with five bumps as described in \cite{pinilla2020}, which were introduced as perturbations in the initial gas surface density profile. They remained static. Specifically, 
    \begin{equation}
        \Sigma^\prime_{\rm{g}}(r)=\Sigma_{\rm{g}}(r)\times\left[1+\sum_i B_i\right],
    \label{Sigma_disk_perturbated}
   \end{equation}
    \noindent with
    \begin{equation}
        B_i(r)=A_i\exp\left(-\frac{(r-r_{\mathrm{p},i})^2}{2w_i^2}\right)
    \label{Gaussian_perturbation}
    .\end{equation}
    The locations ($r_{\mathrm{p},i}$) of these perturbations were at 20, 40, 60, 80, and 100\,au, and the width ($w_i$) of each of them was two local pressure scale heights. The amplitude ($A_i$) for all was 2 (name: smooth static bumps).
        \item As the previous mdoel, but with an amplitude of 4 (name: strong static bumps).
\end{itemize}

The results of these tests are summarized in Fig~\ref{fig:multiple_bumps}, which shows the particle density distribution at 1\,Myr of evolution. When the perturbations were introduced in the gas surface density (Eq.~\ref{Sigma_disk_perturbated}), the bumps were located at the position of the gaps in the other two cases. In addition, for static bumps, the perturbations overlapped in the outer part of the disk. This causes fewer bumps.

After we took these dust density distributions and performed the radiative transfer calculations, the spectral index for the models in which the gaps were introduced in $\alpha$ was 2.8, while the models in which the bumps were introduced in the gas surface density and remained static during the dust evolution was 2.5. The last values are similar to the observed values in different disks in several star-forming regions.

These results show that including more bumps help to decrease the spectral index. The fact that they live from the starting point of the simulations with a given amplitude decreases the values of the spectral index to the averaged values that have been observed. Because the bumps formed by variations in $\alpha$ evolve on viscous timescales,  less trapping is expected than static pressure bumps if these times are longer than
the growth timescale. 
The amplitude here does not seem to play an important role, probably because with this prescription of the bumps and gaps, there is not full filtration, or in other words, radial mixing still allows the grains to move out of the pressure bumps. A deeper investigation is required to confirm this. 

\begin{figure*}
 \centering
        \includegraphics[width=18cm]{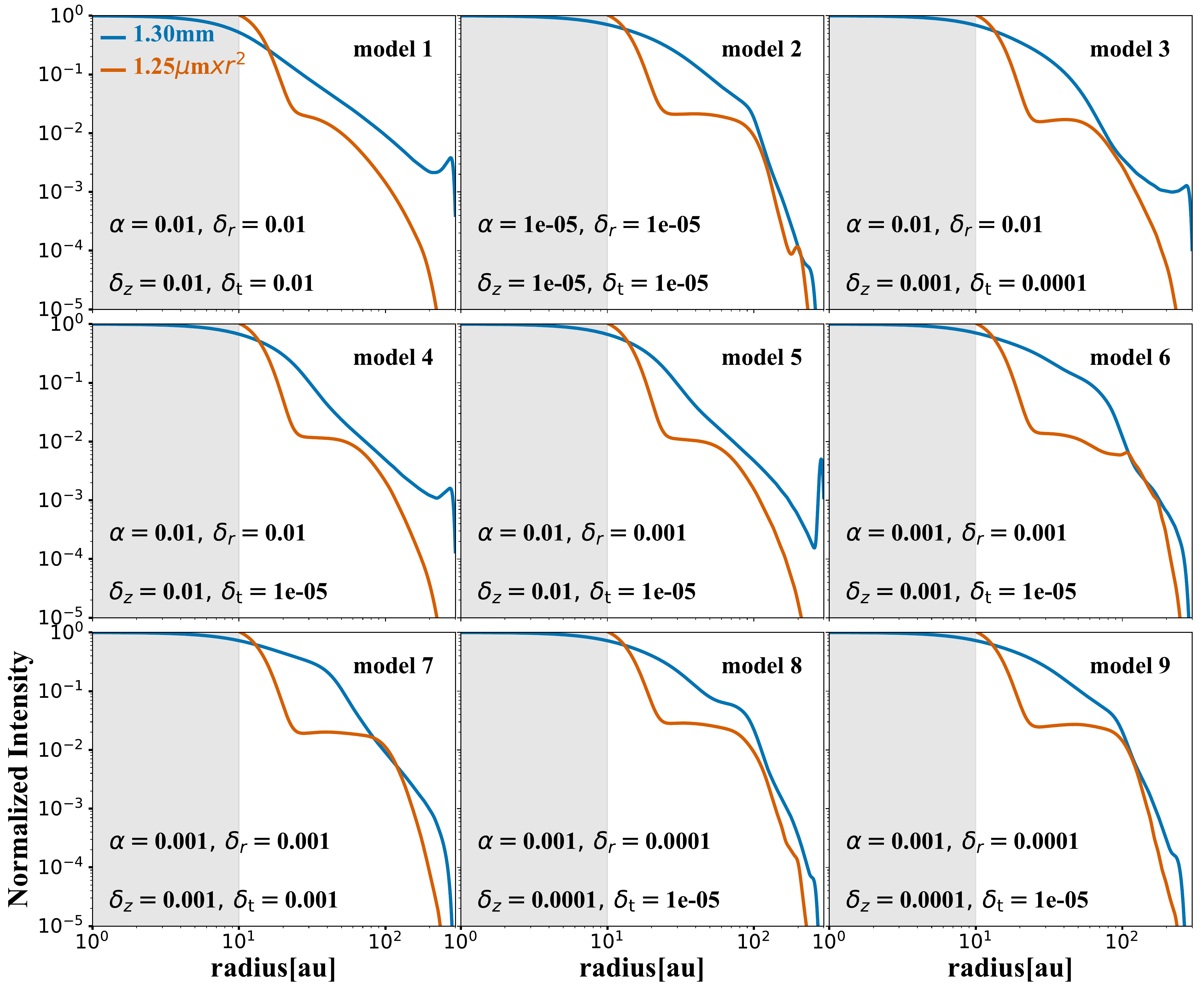}\\
   \caption{Radial intensity profile of the synthetic images (previously convolved with a Gaussian  with an FWHM of 0''.04 in each case) after radiative transfer calculations assuming the particle density distributions from Fig.~\ref{fig:no_bumps_3D}. The intensity profiles are obtained at $\SI{1.25}{\um}$ and 1.3\,mm. The intensity profile at $\SI{1.25}{\um}$ is multiplied by $r^2$ to compensate for the stellar irradiation. The gray regions show the typical size of a coronograph or/and saturated inner pixels from typical observations at scattered light. Case of models without bumps.}
   \label{fig:radial_profile_no_bumps}
\end{figure*}

\subsection{Dust-disk radii} \label{sect:dust_radii}
Using the intensity profile obtained from the synthetic maps at 1.3\,mm, we calculated the radius that enclosed either 68\% ($R_{\mathrm{dust, 68\%}}$) or 90\% ($R_{\mathrm{dust, 90\%}}$) of the emission by assuming a signal-to-noise ratio (S/N) with respect to the maximum emission of 1000 in all cases. We decided to have an S/N compared to the peak instead of a constant S/N for all of the cases because most of the observations available with ALMA have a different sensitivity, all depending on the total flux of each disk and the observation goals.  Currently, this S/N can be obtained for the bright disks with ALMA with a short (some minutes) integration time. This S/N is very challenging to achieve for faint disks, however, and therefore the following discussion is valid when observations with similar S/N with respect to the peak are compared. Tables~\ref{tab:summary_table} and~\ref{tab:summary_table2} include these two dust radii for each simulation. 

Our models without pressure bumps (except for model\,1) span a wide range of values that agree with the values in different star-forming regions, such as Ophiuchus, Taurus, and Lupus \citep{hendler2020}. The mean value of $R_{\mathrm{dust, 68\%}}$ in these models (excluding model\,1) is $\sim53$\,au. In these models, the larger disks are also brighter. Our results demonstrate that a combination of gas and dust diffusion parameters can lead to very different particle density distributions, which directly affects the outer dust radius as obtained from millimeter observations.

Model\,1 is an outlier where the flux is the lowest we found, but this is the largest disk according to our models. This is a consequence of the high turbulence and radial diffusion because the high turbulence produces small grains that are easier to diffuse radially. However, this model has the lowest millimeter flux as well, meaning that we would need deeper observations to detect this outer radius. This result would imply that disks can appear large when observed at millimeter emission without the need of any pressure bump. Recent ALMA observations suggest that large disks remain bright because of substructures, and most of the small disks seem to be faint so far \citep{long2019}. Current observations pose the question of whether the absence of structures is the reason for the small and faint disks. The current data  have some observational bias because deeper observations are needed to determine the outer radius of fainter disks and higher resolution is required to detect structures in the very small disks. 

\begin{figure*}
 \centering
        \includegraphics[width=18cm]{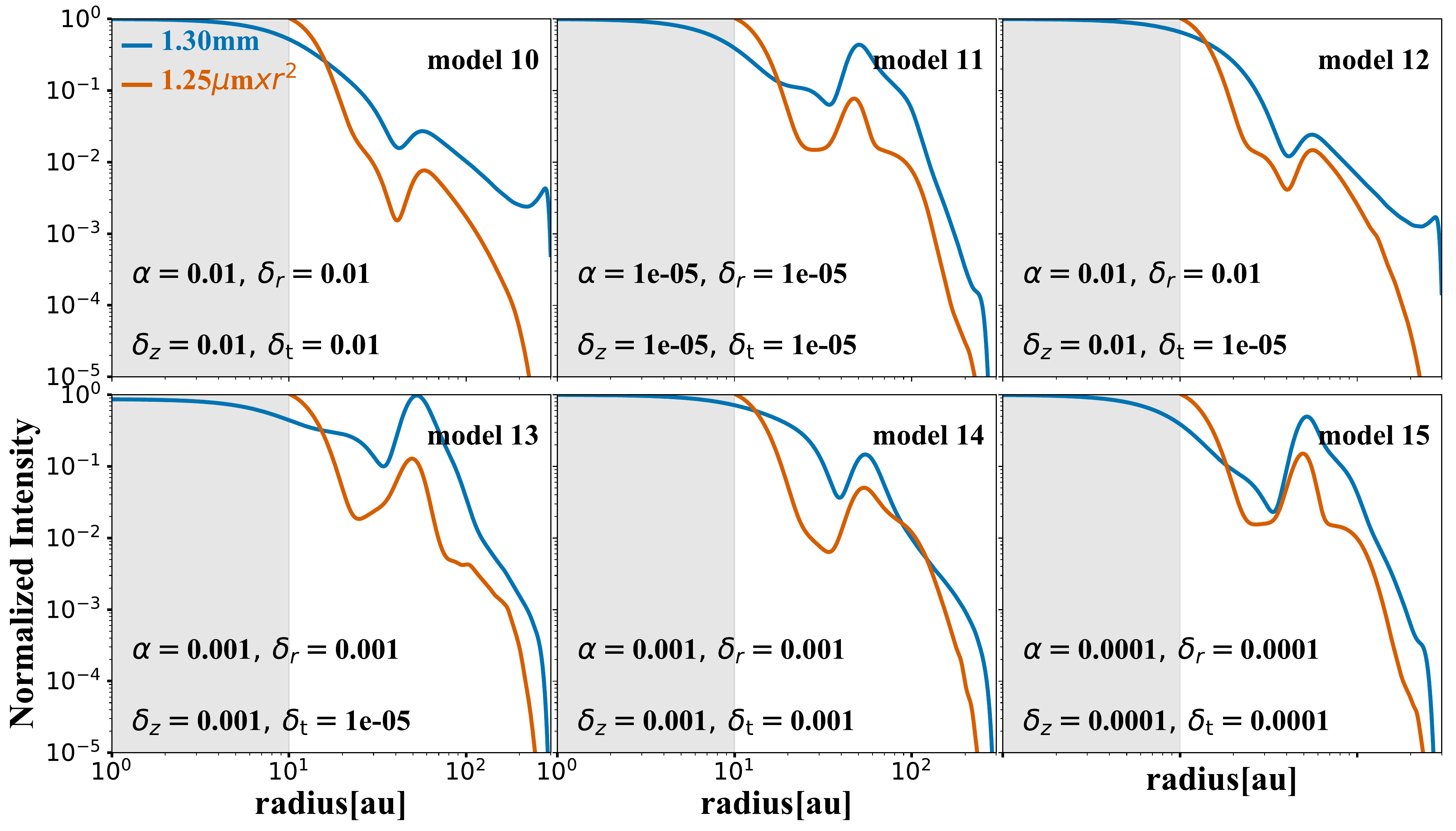}\\
   \caption{As Fig.~\ref{fig:radial_profile_no_bumps}, but when a pressure bump is formed in the outer part of the disk.}
   \label{fig:radial_profile_bumps}
\end{figure*}

The effect of radial diffusion can be more clearly seen when we compare model 4 and model 5, where the only parameter that is different between these two models is $\delta_r$. Model 5 has a $\delta_r$ that is one order of magnitude lower and the disk appears to be smaller, both in the $R_{\mathrm{dust, 68\%}}$ and $R_{\mathrm{dust, 90\%}}$,  with a higher difference when $R_{\mathrm{dust, 90\%}}$  is compared, which is twice as high for the disk with the higher radial diffusion (198\,au for model 4 vs. 95\,au for model 5). In these two models, the flux is very similar, and therefore this result is independent of the S/N of the real observations. 

In our simulations without pressure bumps, $R_{\mathrm{dust, 90\%}}$ can be 1.5 to 3.6 larger than $R_{\mathrm{dust, 68\%}}$, but this strongly depends on the assumption of the S/N. For instance, when we assume half of the S/N of our calculations, $R_{\mathrm{dust, 68\%}}$ remains similar (variations within 5\,au), while $R_{\mathrm{dust, 90\%}}$ decreases such that $R_{\mathrm{dust, 90\%}}$ can be only up twice  $R_{\mathrm{dust, 68\%}}$. 

When models with and without a pressure bump are compared, the dust radii increase for models with a bump, and this increase becomes higher when trapping is efficient (models 11, 13, and 15). The pressure bump is located at $\sim$52\,au in all the models, and the outer radius is larger than this position, either $R_{\mathrm{dust, 68\%}}$ or $R_{\mathrm{dust, 90\%}}$. \cite{pinilla2020} showed that for strong and static pressure bumps, the outer radius traces the location of the farthest pressure bump in the disk. In this case, we do not see this trend, which is probably due to the nature of the pressure bump, which in this work develops with the gas viscous evolution and has a lower amplitude than in the models by \cite{pinilla2020}. The only model in which the outer dust radius does not increase is model 6, which might be the result of the high radial diffusion when compared to the other two cases with trapping. 

\subsection{Substructures}

 Figure~\ref{fig:radial_profile_no_bumps} shows the radial intensity profile of the synthetic images (previously convolved with a Gaussian beam with a full width at half maximum, FWHM, of 0''.04 in each case) after radiative transfer calculations assuming the particle density distributions without any pressure bumps from Fig.~\ref{fig:no_bumps_3D}. The intensity profiles were obtained at $\SI{1.25}{\um}$ and 1.3\,mm. The intensity profile at $\SI{1.25}{\um}$ was multiplied by $r^2$ to compensate for the stellar irradiation. The gray region shows the typical size of a coronograph or/and saturated inner pixels from typical observations at scattered light.

All the radial profiles of the intensity at $\SI{1.25}{\um}$ show a large gap, as reported by \cite{birnstiel2015}, even in the absence of any perturbation in the disk or pressure bump. This gap in scattered light originates from the removal of small grains by sweep-up collisions with the large particles. For this gap to form, there must be an equilibrium between vertical settling/stirring and drift, such that the large grains that are drifting inward are able to sweep-up the small grains in the disk surface. This gap forms in all our models, and the minimum of this gap is located at $\sim25$\,au for all the cases. For most of the simulations, the width of the gap (extend up to $\sim$60-80\,au) and its contrast (a factor of 10) are similar. Model 1 has the narrower and lowest contrast gap, while model\,6 has the widest gap in our simulations (up to $\sim$100\,au). \cite{birnstiel2015} described the location of this gap based on the model assumptions. 

The intensity at 1.3\,mm of the synthetic observations of the models without any pressure bumps decreases smoothly with radius for all the models. In some cases, a local increase in intensity is observed close to the outer radial edge of our simulations. This structure should be taken with caution because it is an effect of assuming a floor value of 10\,K for the temperature profile from $\sim$250\,au. Because of this assumption, the radial profile of the pressure gradient becomes less steep and the large dust particles are locally enhanced. The radial profiles resemble a smooth decrease in the inner parts  together with a sharp decrease in the outer disk. Similar to the dust-disk radii discussed in Sect~\ref{sect:dust_radii}, the location at which the transition from a smooth and sharp profile occurs depends on the gas surface density profile and its vacuous spread outward, the particle size, and on all the dust diffusion parameters that are involved, as discussed in  \cite{birnstiel2014}.

For the models with a bump and in the cases without effective trapping (models 10, 12, and 14), there is a gap in the intensity profile at both wavelengths with a contrast equal to or higher than in scattered light at $\SI{1.25}{\um}$ than in the continuum emission at 1.3\,mm. This is the opposite result: a long-lived and strong trap, from giant planet-disk interaction, where the millimeter gaps have higher contrast \citep[e.g.,][]{ovelar2013, dong2015}. This characteristic, that the gaps have higher contrast in scattered light than at the millimeter, resembles the results from dust-evolution models that include different disk ice-lines \citep{pinilla2017_ice}, where trapping does not occur. Instead, there is a traffic-jam effect that is caused by the changing size of the dust particles at the location of the ice lines, which decreases or increases their radial drift.   The gap in the absence of effective trapping appears to be due to the local decrease in maximum grain size at the gap location. When turbulence sets the maximum grain size (models 10 and 14), the contrast of the scattered light gap (measured as the ratio of the emission of the maximum at the outer edge of the gap and the minimum of the intensity at the gap location) is higher than at millimeter sizes, with a contrast of a factor of 10 in scattered light versus a contrast of a factor of 4-5 at millimeter emission. Without trapping and when the maximum grain size is set by settling (model 12), the contrasts of the gap (a factor of 2) and the width are the same at the two wavelengths.  

For effective trapping (models 11, 13, and 15), all the gaps have higher contrast at  1.3\,mm than at $\SI{1.25}{\um,}$ as expected from previous models \citep[e.g.,][]{ovelar2013}. In model 11, the difference in the contrast between the two wavelengths is lowest, which may have two reasons. First, the large grains formed at pressure maximum, whose opacities are low and that do not significantly contribute to the  1.3\,mm emission. Second, because the gap and bump are introduced as variations in $\alpha$-viscosity and they evolve on a viscous timescale, the gap in model 11 (with $\alpha=10^{-5}$) takes longer to open and reach the final stage. At 1\,Myr of evolution, the bump therefore does not have the same amplitude as in the other cases.

The contrast of the gap is higher in model 15 ($\alpha, \delta_{r,z,\mathrm{t}}=10^{-4}$) than model 13 ($\alpha, \delta_{r,z}=10^{-3}$, and $\TurbPara=10^{-5}$) because the radial diffusion is lower. Lower radial diffusion of the dust helps to keep more grains of different size inside the pressure bump, which contributes to increase the contrast in both wavelengths. 

In all cases, the gap in scattered light seems to be as narrow or wider than at millimeter emission, which is the opposite from models of trapping with a strong and long-lived pressure bump such as those expected from giant planets. The nature of this pressure bump forming with viscous timescales and having lower amplitudes leads to this difference. Therefore the final gap seen in the scattered light in the simulations with a pressure bump may result from a combination of a gap formed purely by dust evolution processes together with the effect of the bump in the distribution of the small dust particles. This characteristic of a wider gap in scattered light versus millimeter emission has been observed in several disks such as TW\,Hya and HD\,163296 \citep[see Fig.~6 in][]{pinilla2017_ice}. However, a deeper investigation that studies how the lifetime of the pressure bumps affects different disk properties is required to test this conclusion.

\section{Summary and conclusions} \label{sect:conclusions}
We investigate the conditions that are required in dust evolution models for growing and trapping pebbles in protoplanetary disks when the fragmentation speed is 1\,m\,s$^{-1}$ in the entire disk. In particular, we separated the parameters controlling the effects of turbulent velocities, vertical stirring, radial diffusion, and viscous evolution of the gas, always assuming that particles cannot diffuse faster (radially or vertically) than the gas. For some of our simulations, we included a smooth gap in the disk by variations of the $\alpha$ viscosity parameter, which led to a pressure bump in the outer disk ($\sim$52\,au). 
The simplified model by \cite{birnstiel2012} does not include the effects of settling-dominated relative particle speeds. Furthermore, for the drift limit they did not consider radial drift-dominated growth. We extended the potential growth limits, including the effects of relative speeds dominated by settling, radial drift, and two different turbulence regimes. The calculations are shown in Appendix~\ref{sec:GrowthBarriers}. We compared our models with observations of protoplanetary disks in near-infrared and millimeter regimes. Our main conclusions are listed below.
\begin{itemize}

\item  With isotropic turbulence, pebble formation can occur in disks when the gas and dust diffusion is low, with values of $10^{-5}-10^{-4}$. Otherwise, pebble formation does not occur, and efficient accumulation of particles in a pressure bump is impossible. 

\item When relative velocities due to turbulence or radial drift are low, settling can dominate the total relative velocities and can set the maximum grain size. In these cases, pebble formation is not possible when $\delta_{z}>10^{-3}$, which also limits the efficiency of trapping when a pressure bump is present. 

\item In general, the millimeter fluxes are lower when pebbles are not formed. In none of the cases did we see a correlation between the processes that set the maximum grain size and the obtained fluxes. All of the fluxes are within the observed values of the disks in different star-forming regions around T-Tauri stars. The fluxes do not change significantly when a trap is included. 

\item Most of the spectral indices in our work are much higher than observed because the number of pebbles is too low as a result of the effective drift. Even for models with one pressure bump and effective trapping, the spectral indices are higher than observed. To solve the discrepancy, multiple bumps assist to decrease $\alpha_{\mathrm{mm}}$. The formation of the bumps and the formation time also affect the spectral index. Static multiple pressure bumps produce lower spectral indices than models in which the gaps and bumps are introduced as variations in $\alpha$-viscosity.

\item Models without pressure bumps span a wide range of synthetic values of $R_{\mathrm{dust, 68\%}}$ that agree with the values in different star-forming regions.  When models with and without a pressure bump were compared, the dust radii increased for models with a bump, and this increased when trapping was efficient.

\item One of our simulations without a pressure bump (model 1, see Table~\ref{tab:summary_table}) produced a large disk as a consequence of the high turbulence, high radial diffusion, and high $\alpha$ that causes the disk to spread and drag particles along. High turbulence limits the formation of pebbles, and small grains are continuously produced by fragmentation; they are easier to diffuse radially. This result would imply that disks can appear large when observed at millimeter emission without the need of any pressure bump. Deeper observations at high angular resolution are needed to test this idea and exclude that substructures are the reason for a potentially large size of a disk. In the case of a bump and effective trapping, high radial diffusion also helps to widen the concentration of large particles in pressure maxima.

\item As found in \cite{birnstiel2015}, even in the absence of any perturbation in the disk or pressure bump, a large gap is formed in the synthetic scattered-light images at $\SI{1.25}{\um}$. For all our models, the minimum of this gap is located around $\sim25$\,au with a similar width (extend up to $\sim$60-80\,au) and contrast (a factor of 10).

\item When a pressure bump is included but there is no effective trapping, a gap opens in the intensity profile at short and long wavelengths, whose contrast is equal to or higher than in scattered light than at millimeter emissions. These results are similar to traffic-jam models, for example, to the dust evolution simulations that include the effect of different disk ice-lines. This is the opposite result of effective trapping by a long-lived and strong trap, for example, from giant planet-disk interaction, where the millimeter gaps have higher contrast.

\end{itemize}

%%%%%%%%%%%%
\section*{Acknowledgements}
We are thankful to the referee for the constructive report. P.P. acknowledges support provided by the Alexander von Humboldt Foundation in the framework of the Sofja Kovalevskaja Award endowed by the Federal Ministry of Education and Research. 
C.T.L. was in parts funded by the Deutsche Forschungsgemeinschaft (DFG,  German  Research  Foundation) as part of the Schwerpunktprogramm (SPP, Priority Program) SPP 1833 ``Building a Habitable Earth''. 
S.M.S. has received funding from the European Research Council (ERC) under the European Union’s Horizon 2020 research and innovation programme under grant agreement No 714769.
S.M.S. acknowledges funding by the Deutsche Forschungsgemeinschaft (DFG, German Research Foundation) Ref no. FOR 2634/1.
We are thankful to A.~Pohl for her insightful help with the coupling of dust evolution models and \textsc{RADMC-3D}. We thank H.~Klahr and R.~Latka for discussions.
%%%%%%%%%%%%%%

\bibliographystyle{aa} % style aa.bst
\bibliography{MAIN.bbl}

\begin{appendix}
\section{Growth barrier estimates}
\label{sec:GrowthBarriers}
In this appendix we provide analytical prescriptions for different growth barriers, which depend on the relative speed among the largest grains. 
We assumed mono-disperse growth for nonvanishing relative velocities for equal-sized particles, and duo-disperse growth where relative velocities of particles of the same size is zero. In the latter case, we model growth assuming that the mass-dominating species grows through collisions of particles that are $N$ times smaller. 
We modeled the relative speed due to radial drift ($\Delta v_\mathrm{r}$), azimuthal drift ($\Delta v_\varphi$), vertical settling ($\Delta v_\mathrm{z}$), and turbulent motion ($\Delta v_\mathrm{turb\,II}$) as
\begin{align}
    \Delta v_\mathrm{r}&\approx\frac{c_\mathrm{s}^2}{v_\mathrm{K}} \left \vert\frac{\partial\ln{P}}{\partial\ln{r}}\right \vert (1-N)\St,
    \\
    \Delta v_\varphi&\approx\frac{1}{2}\frac{c_\mathrm{s}^2}{v_\mathrm{K}} \left \vert\frac{\partial\ln{P}}{\partial\ln{r}}\right \vert\left \vert\frac{1}{\St^2+1}-\frac{1}{N^2\St^2+1}\right \vert,
    \\
    \Delta v_\mathrm{z} &\approx h_\mathrm{d}\Omega(1-N)\St=\cs(1-N)\frac{\St}{\sqrt{1+\St/\delta_{z}}},
    \\
    \Delta v_\mathrm{turb\,II}&\approx\sqrt{3\TurbPara\St}\cs.
\end{align}
These are approximations for the different contributions to the relative speed if $\St<1$  and if the azimuthal pressure gradient is zero. For the second turbulent regime, particles were required to have Stokes numbers larger than $1/\sqrt{\Re}$, where the Reynolds number is given by
\begin{align}
    \nonumber
    \Re&=\frac{\nu_\mathrm{turb}}{\nu_\mathrm{mol}}\approx\frac{\sqrt{\pi}\TurbPara\cs}{\Omega\sqrt{2}\lambda}
    \\
    &=\frac{\sqrt{\pi}\TurbPara n\sigma_{\mathrm{H}_2}\cs}{\Omega}
    \overset{z=0}{\approx}\frac{\TurbPara \sigmag\sigma_{\mathrm{H}_2}}{\sqrt{2} m_\mathrm{g}}.
\end{align}
Here, we adopted the cross section of molecular hydrogen,
\begin{align}
    \sigma_{\mathrm{H}_2}\approx\SI{2E-5}{\cm^2,}
\end{align}
and assumed that the mean molecular weight $m_\mathrm{g}$ is about $2.3$ proton masses, which is roughly
\begin{align}
    m_\mathrm{proton}\approx\SI{1.67E-24}{\g}.
\end{align}
The mean free path of gas molecules following the Maxwell-Boltzmann distribution is given by 
\begin{align}
    \lambda=\frac{1}{\sqrt{2}}\frac{1}{n\sigma_{\mathrm{H}_2}}
\end{align}
with number density of gas molecules $n$. 
We assumed that the collisional behavior can be modeled by collision partners with a factor of $N\approx0.5$ difference in Stokes number (however, see comments at the end of this appendix).  
The factor that depends on the Stokes number in the expression of the the relative azimuthal drift $\Delta v_\varphi$ needs simplifications to find simple analytical expressions for the growth barriers for which the relative speed among the largest grains is dominated by this effect. This expression can be simplified by neglecting higher order terms as follows:
\begin{align}
    \label{eq:phiDriftTermSimpl}
    \left\vert\frac{1}{\St^2+1}-\frac{1}{N^2\St^2+1}\right \vert
    =\frac{(1-N^2)\St^2}{\cancel{N^2\St^4}+(1+N^2)\St^2+1.}
\end{align}
The fragmentation limit can be estimated by setting the dominating relative speed equal to the critical fragmentation speed $v_\mathrm{f}$ \citep{birnstiel2010b,birnstiel2012}. For the case of relative speeds dominated by radial drift, this yields \citep{birnstiel2012}
\begin{align}
    \label{eq:StFragR}
    \St_{\mathrm{frag},r}=\frac{\vfrag\vkep}{\cs^2}\absdlnPdlnr^{-1}\frac{1}{1-N}.
\end{align}
By neglecting the $\mathcal{O}(\St^2)$ term in the denominator on the right side of Eq.~\eqref{eq:phiDriftTermSimpl}, the fragmentation limit in the case of relative speeds dominated by azimuthal drift can be estimated to be
\begin{align}
    \St_{\mathrm{frag},\varphi}=\sqrt{2\frac{\vfrag\vkep}{\cs^2}\absdlnPdlnr^{-1}\frac{1}{1-N^2}}.
\end{align}
However, the relative velocities due to azimuthal drift of the largest grains with similar-sized grains for $\St<1$ are expected to be lower than those caused by relative radial drift: 
\begin{align}
    \label{eq:relvelPhiTooSmall}
    \relvel_\varphi\approx\frac{1-N^2}{2(1-N)}\St\relvel_r<\relvel_r.
\end{align}

For settling-dominated relative speeds, a quadratic equation of the following form is achieved:
\begin{align}
    \St^2-\St\frac{A}{\delta_{z}}-A=0,
\end{align}
where
\begin{align}
    A:=\frac{\vfrag^2}{\cs^2(1-N)^2}.
\end{align}
The physical solution is found to be
\begin{align}
    \label{eq:StFragZ}
    \St_{\mathrm{frag},z}=\frac{A}{2\delta_{z}}+\sqrt{\frac{A^2}{4\delta_{z}^2}+A}.
\end{align}
The fragmentation limit in the turbulence-dominated regime is given by \citep{birnstiel2010,birnstiel2012}
\begin{align}
    \label{eq:StFragTurb}
    \St_{\mathrm{frag,turb}}=\frac{1}{3}\frac{\vfrag^2}{\TurbPara\cs^2}.
\end{align}
Because higher relative velocities lead to critical fragmentation speed at smaller sizes, the overall fragmentation limit can be estimated through
\begin{align}
    \label{eq:fragLim}
    \St_{\mathrm{frag}}=\min{\left\{ \St_{\mathrm{frag,turb}};\,\St_{\mathrm{frag},r};\,\St_{\mathrm{frag},z} \right\}}.
\end{align}

The drift limit can be obtained by setting the growth timescale

\begin{align}
    \tgrow = 2\sqrt{2\pi}\frac{\rhoint\hg}{\sigmad}\frac{a}{\sqrt{1+\St/\DiffZ}\relvel}
\end{align}
and drift timescale
\begin{align}
    \tdrift = \frac{r}{\vert\vdrift\vert}=\frac{\cancel{\St^2}+1}{\St}\frac{r\vkep}{\cs^2}\absdlnPdlnr^{-1} % \overset{\St<1}{\approx}
\end{align}
equal to each other \citep{klahr2006,birnstiel2012}. In the derivation of this growth timescale, we assumed a growth rate (particle radius per time) that is half as high as in canonical derivations from \cite{stepinski1997} and \cite{kornet2001}. When the Smoluchowki equation is used, a factor of 0.5 naturally shows up in the growth rate \citep[e.g.,][]{brauerDissertation2008}. This growth timescale is an estimate for mono-disperse growth only. However, it is still good enough to estimate the growth barriers, which can be fine-tuned with fudge factors or through the N parameter.

With help of the Stokes number in the midplane in the Epstein-drag regime Eq.~\eqref{eq:stokes}, the growth timescale can be written in a form independent of particle radius $a$:
\begin{align}
    \tgrow = 4\sqrt{\frac{2}{\pi}}\frac{\hg}{\relvel}\frac{1}{\Sigdtog}\frac{\St}{\sqrt{1+\St/\DiffZ}},
\end{align}
where $\Sigdtog:=\sigmad/\sigmag$ is the column density ratio of all particles to gas. 
By neglecting second-order terms in the Stokes number, the relative speed due to radial drift can be written as
\begin{align}
    \relvel_r\approx(1-N)\vert\vdrift\vert.
\end{align}
Setting the growth and drift timescales equal leads to the quadratic equation
\begin{align}
    \label{eq:StdriftrQuadraticEq}
    \St^2-\frac{B}{\DiffZ}\St-B=0,
\end{align}
where
\begin{align}
    B:=\left(\frac{\pi(1-N)}{4\sqrt{2\pi}}\frac{\vkep}{\cs}\Sigdtog\right)^2.
\end{align}
The physical solution yields the drift limit in the radial drift-dominated growth regime
\begin{align}
    \label{eq:Stdriftr}
    \St_{\mathrm{drift},r}=\frac{B}{2\DiffZ}+\sqrt{\frac{B^2}{4\DiffZ^2}+B}.
\end{align}

By assuming settling-dominated relative velocities, we obtain
\begin{align}
    \label{eq:StDriftZ}
    \St_{\mathrm{drift},z}=\frac{\sqrt{\pi}}{4\sqrt{2}}(1-N)\Sigdtog\left(\frac{\vkep}{\cs}\right)^2\absdlnPdlnr^{-1}.
\end{align}
The case of dominating relative azimuthal drift is not interesting because of Eq.~\eqref{eq:relvelPhiTooSmall}. 
For turbulent motion as the dominating source of relative velocities, equating the growth and drift timescale after squaring both sides yields
\begin{align}
    \frac{\St^2}{\cancel{\TurbPara/\St}+\TurbPara/\DiffZ}=\sqrt{\frac{3\pi}{32}}\Sigdtog\left(\frac{\vkep}{\cs}\right)^2\absdlnPdlnr^{-1},
\end{align}
which gives (for $\St\gg\DiffZ$)
\begin{align}
    \label{eq:StDriftTurb2}
    \St_\mathrm{drift,turb\,II}=\sqrt{\frac{3\pi}{32}}\sqrt{\frac{\TurbPara}{\DiffZ}}\Sigdtog\left(\frac{\vkep}{\cs}\right)^2\absdlnPdlnr^{-1}.
\end{align}
According to \cite{ormel2007}, particles with $\St<\Re^{-1/2}$ are in the first turbulent regime. This regime can be reached in the very outer parts of circumstellar disks. It has a relative speed
\begin{align}
    \nonumber
    \relvel_{\mathrm{turb\,I}}&\approx\sqrt{\TurbPara}\cs\sqrt{\frac{1-N}{1+N}}
    \\
    \nonumber
    &\hphantom{\approx}\times\underbrace{\sqrt{\frac{\St^2}{\St+\Re^{-1/2}}-\frac{N^2\St^2}{N\St+\Re^{-1/2}}}}_{\approx\sqrt{(1-N^2)}\Re^{1/4}\St}
    \\
    &\approx\sqrt{\TurbPara}\cs(1-N)\Re^{1/4}\St.
\end{align}
Following the same steps as above to obtain the drift limit, the same quadratic equation as Eq.~\eqref{eq:StdriftrQuadraticEq} with the solution Eq.~\eqref{eq:Stdriftr} is obtained, but for $B\rightarrow C$ with
\begin{align}
    C:=\frac{\pi}{32}\left(\frac{\vkep}{\cs}\right)^4\Sigdtog^2\absdlnPdlnr^{-2}\TurbPara(1-N)^2\sqrt{\Re}.
\end{align}
The solution is given by
\begin{align}
    \label{eq:StdriftTurb1}
    \St_{\mathrm{drift,turb\,I}}=\frac{C}{2\DiffZ}+\sqrt{\frac{C^2}{4\DiffZ^2}+C}.
\end{align}
Because of the different scaling of the relative velocities, this drift limit $\St_{\mathrm{drift,turb\,I}}$ fits the simulation results better for $N=0.1$. The overall drift limit in the turbulence-dominated case is estimated as
\begin{align}
    \St_{\mathrm{drift,turb}}=
    \begin{cases}
        \St_{\mathrm{drift,turb\,II}} & ,\,\St_{\mathrm{drift,turb\,I}}\geq\Re^{-1/2}
        \\
        \St_{\mathrm{drift,turb\,I}}  & ,\,\St_{\mathrm{drift,turb\,I}}<\Re^{-1/2}
    \end{cases}
    .
\end{align}

Because faster growth caused by higher relative velocities leads to shorter growth timescales, the overall drift limit is given by
\begin{align}
    \label{eq:driftLim}
    \St_{\mathrm{drift}}=\max{\left\{ \St_{\mathrm{drift,turb}};\,\St_{\mathrm{drift},r};\,\St_{\mathrm{drift},z} \right\}}.
\end{align}
Both the fragmentation limit Eq.~\eqref{eq:fragLim} and the drift limit Eq.~\eqref{eq:driftLim} can be converted into particle radii using the $\St$-$a$ relation in the Epstein regime Eq.~\eqref{eq:stokes}. 

The $N$ parameter can potentially be different for the fragmentation limit or drift limit because this depends on the size distribution, which is different in these limits.
While the size distribution in drift-limited regions is very top heavy, fragmentation-limited regions show shallower size distributions. 
In the most cases, $N=0.3$ for fragmentation limits and $N=0.8$ for drift limits mimic the evolution of the maximum grain size quite well. 
We summarize the values used for $N$ in different cases in Tab.~\ref{tab:paramN}. 
Especially the fragmentation limits work very well. The drift limit at late times and in the most outer regions of the disk can significantly deviate from the data of numerical simulations. This could be an effect of violated conditions for the analytical estimates (e.g., relative speeds are not dominated by one effect, or the size distribution does not allow an almost mono-disperse or duo-disperse approach).

\begin{table}[t]
\caption{Parameter values of the relative Stokes number $N$ between favorable collision partners used to model the maximum size in different regimes of relative speed. 
} 
\label{tab:paramN}
\centering
\begin{tabular}{lcc}
 \toprule\toprule
 \\
 $\relvel$ regime & $N$ in drift limit & $N$ in frag. limit
 \\
 \midrule
 turb I           & 0.1, Eq.~\eqref{eq:StdriftTurb1}     & --
 \\
 turb II          & 1.0, Eq.~\eqref{eq:StDriftTurb2}     & 1.0, Eq.~\eqref{eq:StFragTurb}
 \\
 rad. drift       & 0.8, Eq.~\eqref{eq:Stdriftr}         & 0.3, Eq.~\eqref{eq:StFragR}
 \\
 settling         & 0.8, Eq.~\eqref{eq:StDriftZ}         & 0.3, Eq.~\eqref{eq:StFragZ}
 \\
 \bottomrule
 %\bottomrule
\end{tabular}
\end{table}

\end{appendix}

\end{document}